\documentclass[12pt]{article}
\usepackage[a4paper]{geometry}
\usepackage[english]{babel}

\usepackage{amsmath}
\usepackage{amsfonts}
\usepackage{amstext}
\usepackage{amssymb}
\usepackage{amsthm}
\usepackage{amscd}

\usepackage[pagebackref,draft=false]{hyperref}
\hypersetup{colorlinks,
linkcolor=myrefcolor,
citecolor=mycitecolor,
urlcolor=myurlcolor}

\usepackage[capitalize]{cleveref}
\usepackage{caption}

\usepackage{xcolor}
\definecolor{myurlcolor}{rgb}{0,0,0.4}
\definecolor{mycitecolor}{rgb}{0,0.5,0}
\definecolor{myrefcolor}{rgb}{0.5,0,0}
\usepackage{graphicx}
\usepackage{tikz}
\usepackage{tikz-cd}

\usepackage{etoolbox}
\usepackage{makeidx}
\usepackage{authblk} 
\usepackage{sectsty}
\usepackage{mathrsfs}
\usepackage{dsfont}
\usepackage{enumitem} 
\usepackage[]{latexsym}
\usepackage{braket}
\usepackage{caption}

\newtheorem{remark}{Remark}

\newtheorem{theorem}{Theorem}

\newtheorem{proposition}{Proposition}
\newtheorem{definition}{Definition}

\newtheorem*{proof*}{Proof}

\newcommand{\be}{\begin{equation}}
\newcommand{\ee}{\end{equation}}
\newcommand{\bea}{\begin{eqnarray}}
\newcommand{\eea}{\end{eqnarray}}
\newcommand{\vsp}{\vspace{0.4cm}}
\newcommand{\grit}[1]{{\bfseries {\itshape {#1}}}}

\newcommand{\obsp}{\mathfrak{O}}
\newcommand{\stsp}{\mathscr{S}}

\newcommand{\Uh}{\mathcal{U}(\mathcal{H})}
\newcommand{\uh}{\mathfrak{u}(\mathcal{H})}

\newcommand{\SUh}{\mathcal{SU}(\mathcal{H})}

\newcommand{\suh}{\mathfrak{su}(\mathcal{H})}
\newcommand{\sustarh}{\mathfrak{su}^{*}(\mathcal{H})}

\newcommand{\traceh}{\mathfrak{T}^{1}(\mathcal{H})}

\newcommand{\gr}{\mathrm{g}}

\newcommand{\SLh}{\mathcal{SL}(\mathcal{H})}
\newcommand{\Hh}{\mathcal{H}}
\newcommand{\stsph}{\mathscr{S}(\mathcal{H})}

\newcommand{\Bh}{\mathcal{B}(\mathcal{H})}
\newcommand{\slh}{\mathfrak{sl}(\mathcal{H})}

\newcommand{\Tr}{\textit{Tr}}

\author[1,7]{D. Chru\'sci\'nski}
\author[2,3,8]{F. M. Ciaglia}
\author[5,6,9]{A. Ibort}
\author[3,4,10]{G. Marmo}
\author[3,4,11]{F. Ventriglia}

\affil[1]{\textit{\footnotesize Institute of Physics, Faculty of Physics, Astronomy and Informatics
Nicolaus Copernicus University, Grudzi\c{a}dzka 5/7, 87–100 Toru\'{n}, Poland.}}

\affil[2]{\textit{\footnotesize Max Planck Institute for Mathematics in the Sciences, Inselstrasse 22, 04103 Leipzig, Germany}}

\affil[3]{\textit{\footnotesize INFN-Sezione di Napoli, Complesso Universitario di Monte S. Angelo Edificio 6, via Cintia, 80126 Napoli, Italy.}}

\affil[4]{\textit{\footnotesize Dipartimento di Fisica ``E. Pancini'', Universit\`a di Napoli Federico II, Complesso Universitario di Monte S. Angelo Edificio 6, via Cintia, 80126 Napoli, Italy.}}

\affil[5]{\textit{\footnotesize ICMAT, Instituto de Ciencias Matem\'{a}ticas (CSIC-UAM-UC3M-UCM).}}

\affil[6]{\textit{\footnotesize Depto. de Matem\'aticas, Univ. Carlos III de Madrid, Avda. de la Universidad 30, 28911 Legan\'es, Madrid, Spain.}}

\affil[7]{\footnotesize e-mail: \texttt{darch@fizyka.umk.pl}}

\affil[8]{\footnotesize e-mail: \texttt{florio.ciaglia@mis.mpg.de}, \texttt{florio.m.ciaglia@gmail.com}}

\affil[9]{\footnotesize e-mail: \texttt{marmo@na.infn.it}} 

\affil[10]{\footnotesize e-mail: \texttt{albertoi@math.uc3m.es}}

\affil[11]{\footnotesize e-mail: \texttt{ventriglia@na.infn.it}}

\date{}

\title{Stratified Manifold of Quantum States, actions of the complex special linear group}

\begin{document}

\maketitle

\tableofcontents

\vsp
\vsp
\vsp
\vsp

\vsp
\vsp

\abstract{We review the geometry of the space of quantum states $\stsph$ of a finite-level quantum system with Hilbert space $\Hh$ from a group-theoretical point of view.
This space carries two stratifications generated by the action of two different Lie groups, namely, the special unitary group $\SUh$ and its complexification $\SLh$, the complex special linear group.
A stratum of the stratification generated by $\SUh$ is  composed of isospectral states, that is, density operators with the same spectrum,
A stratum of the stratification generated by $\SLh$ is composed of quantum states with the same rank.

We prove that on every submanifold of isospectral quantum states there is also a canonical left action of $\SLh$ which is related with the canonical K\"{a}hler structure on isospectral quantum states.
The fundamental vector fields of this $\SLh$-action are divided into Hamiltonian and gradient vector fields.
The former give rise to invertible maps on $\stsph$ that preserve the von Neumann entropy and the convex structure of $\stsph$, while the latter give rise to invertible maps on $\stsph$ that preserve the von Neumann entropy but not the convex structure of $\stsph$.

A similar decomposition is given for the fundamental vector fields of the $\SLh$-action generating the stratification of $\stsph$ into manifolds of quantum states with the same rank.
However, in this case, the gradient vector fields preserve the rank but do not preserve entropy.

Finally, some comments on multipartite quantum systems are made, and it is proved that the sets of product states of a multipartite quantum system are homogeneous manifolds for the local action of the complex special linear group associated with the partition.}

\section{Introduction}\label{ch: Introduction}

Our understanding of the geometry of the space of quantum states is in constant evolution and there are different fields of application in which it is possible to use the knowledge we gain.
For instance, geometrical ideas have been successfully exploited when addressing the foundations of quantum mechanics \cite{ashtekar_schilling-geometrical_formulation_of_quantum_mechanics, cantoni-generalized_transition_probability, cantoni-the_riemannian_structure_on_the_space_of_quantum-like_systems, carinena_clemente-gallardo_marmo-geometrization_of_quantum_mechanics, chruscinski_marmo-remarks_on_the_gns_representation_and_the_geometry_of_quantum_states, ciaglia_ibort_marmo-geometrical_structures_for_classical_and_quantum_probability_spaces, cirelli_lanzavecchia_mania-normal_pure_states_and_the_von_neumann_algebra_of_bounded_operators_as_kahler_manifold, cirelli_mania_pizzocchero-quantum_mechanics_as_an_infinite_dimensional_Hamiltonian_system_with_uncertainty_structure,  ercolessi_marmo_morandi-from_the_equations_of_motion_to_the_canonical_commutation_relations, grabowski_kus_marmo-on_the_relation_between_states_and_maps_in_infinite_dimensions, kibble-geometrization_of_quantum_mechanics, marmo_scolarici_simoni_ventriglia-the_quantum-classical_transition:the_fate_of_the_complex_structure}, quantum information theory \cite{ay_tuschmann-duality_versus_flatness_in_quantum_information_geometry,  ciaglia_dicosmo_felice_mancini_marmo_perez-pardo-aspects_of_geodesical_motion_with_fisher-rao_metric:classical_and_quantum, ciaglia_dicosmo_laudato_marmo_mele_ventriglia_vitale-a_pedagogical_intrinsic_approach_to_relative_entropies_as_potential_functions_of_quantum_metrics, facchi_kulkarni_manko_marmo_sudarshan_ventriglia-classical_and_quantum_fisher_information_in_the_geometrical_formulation_of_quantum_mechanics, laudato_marmo_mele_ventriglia_vitale-tomographic_reconstruction_of_quantum_metrics, manko_marmo_ventriglia_vitale-metric_on_the_space_of-quantum_states_from_relative_entropy_tomographic_reconstruction, cencov_morozowa-markov_invariant_geometry_on_state_manifolds, petz-monotone_metrics_on_matrix_spaces, wootters-statistical_distance_and_hilbert_space}, quantum dynamics \cite{carinena_clemente-gallardo_jover-galtier_marmo-tensorial_dynamics_on_the_space_of_quantum_states, chruscinski_jamiolkowski-geometric_phases_in_classical_and_quantum_mechanics, chruscinski_pascazio-a_brief_history_of_the_gkls_equation, ciaglia_dicosmo_ibort_laudato_marmo-dynamical_vector_fields_on_the_manifold_of_quantum_states, ciaglia_dicosmo_ibort__marmo-dynamical_aspects_in_the_quantizer-dequantizer_formalism, ciaglia_dicosmo_laudato_marmo-differential_calculus_on_manifolds_with_boundary.applications,  cirelli_pizzocchero-on_the_integrability_of_quantum_mechanics_as_an_infinite_dimensional_system}, entanglement theory \cite{aniello_clemente-gallardo_marmo_volkert-classical_tensors_and_quantum_entanglement_I, bengtsson_zyczkowski-geometry_of_quantum_states:_an_introduction_to_quantum_entanglement, chirco_mele_oriti_vitale-fisher_metri_geometric_entanglement_and_spin_networks, grabowski_kus_marmo-geometry_of_quantum_systems_density_states_and_entanglement, grabowski_kus_marmo-symmetries_group_actions_and_entanglement, huckleberry_kus_sawicki-bipartite_entanglement_spherical_ations_and_geometry_of_local_unitary_orbits, huckleberry_kus_sawicki-symplectic_geometry_of_entanglement, kus_oszmaniec_sawicki-convexity_of_momentum_map_morse_index_and_quantum_entanglement}.

The purpose of this paper is to present a review of well-known and lesser-known ideas, facts and constructions regarding the geometry of the space $\stsph$ of a quantum system with Hilbert space $\Hh$.
In order to avoid the technicality of infinite-dimensional differential geometry, we will consider finite level quantum systems for which $\mathrm{dim}(\Hh)$ is finite.

\vsp

We start with the space $\stsp_{1}(\Hh)$ of pure quantum states, that is, the extremal points of the convex set $\stsph$ of quantum states. 
For  a quantum system modelled on the Hilbert space $\Hh$,  the space $\stsp_{1}(\Hh)$  may be identified with the complex projective space $\mathbb{CP}(\Hh)$ associated with $\Hh$.
It is well-known that the complex projective space $\mathbb{CP}(\Hh)$ is a homogeneous space for the special unitary group $\SUh$, and this allows for the definition of a canonical K\"{a}hler structure on it (also in infinite-dimensions).
What is lesser-known is that $\mathbb{CP}(\Hh)$ is also a homogeneous space of the complex special linear group $\SLh$, that is, the complexification of $\SUh$.
The action of $\SLh$ on $\mathbb{CP}(\Hh)$ ``comes from'' the canonical linear action of $\mathcal{GL}(\Hh)$ on $\Hh$, which is projectable on $\mathbb{CP}(\Hh)$ and ``becomes'' an action of $\SLh$.
However when additional structures are considered on $\mathbb{CP}(\Hh)$ (e.g., the K\"{a}hler structure) this group is reduced to the subgroup of unitary transformations.

As said before, the space $\stsp_{1}\cong\mathbb{CP}(\Hh)$ of pure quantum states is a compact K\"{a}hler manifold \cite{ashtekar_schilling-geometrical_formulation_of_quantum_mechanics, carinena_clemente-gallardo_marmo-geometrization_of_quantum_mechanics, chruscinski_marmo-remarks_on_the_gns_representation_and_the_geometry_of_quantum_states, cirelli_lanzavecchia_mania-normal_pure_states_and_the_von_neumann_algebra_of_bounded_operators_as_kahler_manifold, cirelli_mania_pizzocchero-quantum_mechanics_as_an_infinite_dimensional_Hamiltonian_system_with_uncertainty_structure, ercolessi_marmo_morandi-from_the_equations_of_motion_to_the_canonical_commutation_relations, kibble-geometrization_of_quantum_mechanics}, and thus, it admits a symplectic form $\omega$, a Riemannian metric $g$ and complex structure $J$ such that $g=\omega\circ J$.
All these tensors are invariant with respect to the canonical action of $\SUh$ on $\mathbb{CP}(\Hh)$.
However, it should be noticed that, while it preserves the complex structure $J$, the action of the complex special linear group $\SLh$ takes us from one K\"{a}hler structure to a different one, unless we restrict the action to the subgroup of symplectic transformations, in which case they turn out to be also unitary because they must preserve the complex structure.
By preserving the complex structure, the factorization by means of the symplectic structure and the symmetric structure changes.
Thus, we obtain alternative K\"{a}hlerian structures all of them associated with the same complex structure.
Consequently, we may get alternative realizations of the Unitary group.
As the action is not linear the linear structure will be changed.

For a finite-level quantum system we may identify the dual $\sustarh$ of the Lie algebra $\suh$ with the space of linear Hermitean operators on $\Hh$ with trace equal to $1$, and, by considering the moment map $\mu\colon \mathbb{CP}(\Hh)\rightarrow\suh^{*}$ associated with the canonical symplectic action of $\SUh$ on $\mathbb{CP}(\Hh)$, we may build the convex hull $\stsp$ of $\mu(\mathbb{CP}(\Hh))$ in the space of linear Hermitean operators on $\Hh$.
This convex hull $\stsph$ is the space of quantum states of the system  and it is identified with the space of density operators, that is, Hermitean linear operators on $\Hh$ that are positive semi-definite and have trace equal to $1$.
In this setting, the image $\mu(\mathbb{CP}(\Hh))$ of the space of pure quantum states is the space of rank-one Hermitean projectors.
The Lie group $\SUh$ acts on $\suh^{*}$ by means of the coadjoint action and it is possible to prove that this coadjoint action preserves $\stsph$.
In particular, we obtain a partition of the space of quantum states $\stsph$ into  the disjoint union of coadjoint orbits of $\SUh$.
These orbits are labelled by the spectrum of the quantum states they contain (up to permutations), and are endowed with the structure of K\"{a}hler manifolds.

As the space of mixed quantum states is obtained from convex combinations of pure quantum states, it is a natural question to ask what happens of the action of the special linear group.
It has been argued elsewhere that even though the action of the $\SLh$ group does not preserve the metric structure, it preserves   Hermiticity, positivity and rank of the states.
Recently, it has been proved that there is a (nonlinear) action of the special linear group $\SLh$ on the whole $\stsph$ \cite{ciaglia_dicosmo_ibort_laudato_marmo-dynamical_vector_fields_on_the_manifold_of_quantum_states, ciaglia_ibort_marmo-differential_geometry_of_quantum_states_observables_and_evolution, grabowski_kus_marmo-geometry_of_quantum_systems_density_states_and_entanglement}.
This Lie group is the complexification of the special unitary group and its action on $\stsph$ allows us to build the submanifolds of quantum states with fixed rank.
These submanifolds are ``thicker'' than the manifolds of isospectral quantum states, except for the case of pure quantum states which is a homogeneous space for both $\SUh$ and $\SLh$.
Unfortunately, the manifolds of quantum states with fixed rank do not carry ``obvious'' additional geometrical structures other than the differential structure and the smooth action of $\SLh$.
However, exploiting the $C^{*}$-algebra structure of $\mathcal{B}(\Hh)$, or, equivalently, the Jordan-Lie algebra structure of the space of Hermitean linear operators on $\Hh$, it is possible to introduce two contravariant bivector fields $\Lambda$ and $\mathcal{R}$ on the affine hyperplane $\mathfrak{T}_{1}(\Hh)$ of Hermitean operators with trace equal to $1$ by means of which it is possible to describe the infinitesimal action of $\SLh$ on $\stsph$ using Hamiltonian vector fields associated with $\Lambda$ and gradient-like vector fields associated with $\mathcal{R}$.

\vsp

In sections \ref{sec: Quantum states and the special unitary group} and \ref{sec: Kahler structures on the manifolds of isospectral quantum states} we review the differential geometry of the actions of $\SUh$ and $\SLh$ on $\stsp$, and the K\"{a}hler structure of the manifolds of isospectral quantum states (orbits of $\SUh$).
Then, in section \ref{sec: q-bit I}, we explicitly work out the form of these actions in the case of a two-level quantum system, i.e., the q-bit.
In section \ref{sec: Lie groups, Kahler actions, and complexification} we present our main result and in section \ref{sec: composite systems} we comment on its implication in the case of composite systems.
Specifically, we prove that the special linear group $\SLh$ acts not only on the manifolds of quantum states with fixed rank, but also on the manifolds of isospectral quantum states (orbits of $\SUh$), and, in the latter case, this action is associated with Hamiltonian and gradient vector fields.
The former give rise to invertible maps on $\stsp$ that preserve the von Neumann entropy and the convex structure of $\stsp$, while the latter give rise to invertible maps on $\stsp$ that preserve the von Neumann entropy but not the convex structure of $\stsp$.
We prove this result in the slightly more general context of a K\"{a}hler manifold on which there is a Lie group $G$ acting by means of automorphisms for the K\"{a}hler structure.

Section \ref{sec: q-bit II} allows us to explicitly work out the details of our main result in the case of a q-bit  making everything less abstract and more concrete.

Section \ref{sec: composite systems} deals with bipartite quantum systems of distinguishable ``particles'' with $\Hh=\Hh_{A}\otimes\Hh_{B}$.
Specifically, we analyze separability properties of mixed quantum states with respect to the local action of the special linear group $\mathcal{SL}(\Hh_{A})\times\mathcal{SL}(\Hh_{B})$  preserving the rank.
We introduce the notion of $(A,B)$-rank for quantum states and we prove that the spaces of quantum states with fixed $(A,B)$-rank are smooth manifolds that are homogeneous spaces of $\mathcal{SL}(\Hh_{A})\times\mathcal{SL}(\Hh_{B})$.
In particular, the manifold of pure quantum states and the manifold of invertible product states turn out to be explicit examples of spaces quantum states with fixed $(A,B)$-rank.

\section{Geometry of quantum states}
\label{sec: Quantum states and the special unitary group}

Let us consider a quantum system with Hilbert space $\Hh$ such that $\mathrm{dim}(\Hh)<+\infty$.
Denoting by $\Bh$ the space of bounded linear operators on $\Hh$, the space of quantum observables $\obsp$ is the space of Hermitean linear operators in $\Bh$.
This is a real vector space of which $\Bh$ may be thought of as being the complexification.
In this setting, the space of quantum states $\stsph$ is the space of density operators, that is:

\be
\stsph:=\left\{\rho\in\obsp\colon  \Tr(\rho)=1,\;\;\;\rho \mbox{ is positive }\right\}\,.
\ee
The normalization imposed by the constraint on the trace allows us to look at  $\stsph$ as a convex body in the affine hyperplane:

\be
\mathfrak{T}_{1}(\Hh):=\left\{\xi\in\obsp\colon \Tr(\xi)=1\right\}\,.
\ee
%

\begin{remark}\label{rem: lower bound on purity depends on rank}

Let $\{\tau^{j}\}_{j=1,...,(n^{2}-1)}$ with be a basis in $\suh$ such that:

\be
\mathrm{Tr}\left(\tau^{j}\,\tau^{k}\right)\,=\,-\delta^{jk}\,,
\ee
and set $\tau^{0}=\frac{\mathbb{I}}{\sqrt{n}}$, where $\mathbb{I}$ is the identity operator on $\Hh$.
Let $c^{jk}_{l}$ be the structure constants  of $[\cdot,\cdot]$ relative to  $\{\tau^{j}\}_{j=1,...,(n^{2}-1)}$, that is:

\be
[\tau^{j},\,\tau^{k}]=\,c^{jk}_{l}\,\tau^{l}\,,
\ee
and let the coefficients $d^{jk}_{l}$ and $d^{jk}_{0}$ be defined according to:

\be
\tau^{j}\odot\tau^{k}\,=\,-\imath\left(\tau^{j}\,\tau^{k} + \tau^{k}\,\tau^{j}\right)\,=\, d^{jk}_{l}\tau^{l} - 2\,\delta^{jk}\tau^{0}\,.
\ee
Denote with $\{x^{j}\}_{j=1,...,(n^{2}-1)}$ the  coordinate system   on $\mathfrak{T}_{1}(\Hh)$ associated with $\{\tau^{j}\}_{j=1,...,(n^{2}-1)}$ by means of:

\be
x^{j}(\xi):=-\imath\Tr\left(\tau^{j}\,\xi\right)\,,
\ee
so that every $\xi\in\mathfrak{T}_{1}(\Hh)$ may be written as:

\be
\xi\,=\,\frac{1}{\sqrt{n}}\,\varsigma_{0} + x^{j}\,\varsigma_{j}\,,
\ee
with $\varsigma_{0}=-\imath\tau^{0}$ and $\varsigma_{j}\,=\,-\imath\tau^{j}$ Hermitean elements in $\Bh$.

A quantum state $\rho$ is an element in $\mathfrak{T}_{1}(\Hh)$ such that $\rho\geq0$ and $\mathrm{Tr}(\rho^{2})\leq 1$ where the equality holds if and only if $\rho$ is pure.
Now, the purity function $\mathrm{Tr}(\rho^{2})$ may be written as:

\be
\mathrm{Tr}(\rho^{2})\,=\,\frac{1}{n} + \delta_{jk}\,x^{j}\,x^{k}\,,
\ee
and thus:

\be
\mathrm{r}^{2}(\rho)\,:=\,\mathrm{Tr}(\rho^{2}) - \frac{1}{n^{2}}\,=\,\delta_{jk}\,x^{j}\,x^{k}\,\leq \frac{n -1}{n}
\ee
for every quantum state $\rho\in\mathfrak{T}_{1}(\Hh)$.
Clearly, $\mathrm{r}^{2}(\rho)$ attains its minimum $\frac{1}{n}$ at the maximally mixed quantum state $\rho_{m}=\frac{\mathbb{I}}{n}$.

On the other hand, $\mathrm{Tr}(\rho^{2})$ may be written in terms of the eigenvalues $\lambda^{j}$ of $\rho$:

\be
\mathrm{Tr}(\rho^{2})\,=\,\sum_{j=1}^{k}\,\left(\lambda^{j}\right)^{2}
\ee
where $k$ is the rank of $\rho$ as a linear operator on $\Hh$.
Consequently, given $\rho$, we may build a ``virtual'' quantum state $\rho_{k}$  on a Hilbert space $\Hh_{k}$ (with $\mathrm{dim}(\Hh_{k})=k)$ by selecting an orthonormal basis in $\Hh_{k}$ and declaring $\rho_{k}$ to be diagonal in this basis with eigenvalues equal to the $\lambda^{j}$'s.
Conseqently, we obtain:

\be
\mathrm{Tr}(\rho^{2})\,=\,\mathrm{Tr}(\rho_{k}^{2})\geq \frac{1}{k}
\ee
because $\mathrm{Tr}(\rho_{k}^{2})$ is bounded from below by its value on the maximally mixed state on $\Hh_{k}$.
It is then clear that a quantum state in $\mathfrak{T}_{1}(\Hh)$ with rank $k$ is such that:

\be
\begin{split}
\frac{1}{k}\,&\leq\,\mathrm{Tr}(\rho^{2})\,\leq\,1 \\
\frac{n-k}{nk}\,&\leq\,\mathrm{r}^{2}(\rho)\,\leq\, \frac{n -1}{n}\,.
\end{split}
\ee
The lower bounds on $\mathrm{r}^{2}(\rho)$ will play a role in section \ref{sec: composite systems} when dealing with composite systems.
Specifically, there are quantum states that are separable with respect to any decomposition of the Hilbert space \cite{gurvits_barnum-separable_balls_around_the_maximally_mixed_multipartite_quantum_state, zyczkowski_horodecki_sanpera_lewenstein-volume_of_the_set_of_separable_states}, and we will see that all these separable states must have maximal rank precisely because of the lower bound on $\mathrm{r}^{2}(\rho)$.
\end{remark}

Consider now the unitary group $\Uh$ and the special unitary group $\SUh$, and denote by $\uh$ and $\suh$, respectively, their Lie algebras.
It is well known that the Lie algebra $\uh$ of $\Uh$ is isomorphic with the the real vector space of skew-Hermitean linear operators on $\Hh$  and that the Lie algebra $\mathfrak{su}(\mathcal{H})$ consequently would be isomorphic with the subspace of traceless skew-Hermitean linear operators on $\mathcal{H}$.
Every element $\mathbf{A}$ in $\mathfrak{u}(\mathcal{H})$ may be written as $\imath\mathbf{a}$ where $\mathbf{a}$ is Hermitean, in particular, if $\Tr(\mathbf{a})=0$ then $\mathbf{A}$ is in $\suh$.
There is a faithful left action of $\SUh$ on $\mathfrak{T}_{1}(\Hh)$, and thus on the space $\stsph$ of quantum states, given by:
 
\be\label{eqn: action of SUh on quantum states}
\xi\;\;\mapsto\;\;\mathbf{U}\,\xi\,\mathbf{U}^{\dagger}\,,
\ee
where $\xi\in\mathfrak{T}_{1}(\Hh)$, and $\mathbf{U}$ is an element of the special unitary group $\SUh$.
This action is linear, and thus preserve the convex structure of $\stsph$.
Furthermore, because  $\mathbf{U}^{\dagger}=\mathbf{U}^{-1}$, it is a similarity transformation, which means that it preserves the spectrum of $\xi\in\mathfrak{T}_{1}(\Hh)$.
The orbits of this action lying in $\stsph$  are known as the submanifolds of isospectral quantum states.

The Lie algebra structure of $\uh$ allows us to define a  bivector field $\Lambda$ on $\mathfrak{T}_{1}(\Hh)$.
In order to do so, we note that every $\mathbf{A}$ in $\mathfrak{u}(\mathcal{H})$ defines a linear function $f_{\mathbf{A}}\colon\mathfrak{T}_{1}(\mathcal{H})\rightarrow\mathbb{R}$ as follows:

\be\label{eqn: expectation value functions}
f_{\mathbf{A}}(\xi):=-\imath\Tr\left(\mathbf{A}\,\xi\right)\,,
\ee
and, since the exterior differentials of the functions $f_{\mathbf{A}}$ with $\mathbf{A}\in\uh$ generate the cotangent space at each point of $\mathfrak{T}_{1}(\Hh)$, the bivector field $\Lambda$ is uniquely determined by setting:

\be
\Lambda(\mathrm{d}f_{\mathbf{A}},\,\mathrm{d}f_{\mathbf{B}}):=f_{[\mathbf{A},\mathbf{B}]}\,,
\ee
where $[\cdot,\cdot]$ is the Lie product in $\uh$, and extending it by linearity.
The Jacobi identity for the Lie product $[\cdot,\cdot]$ implies that the Schouten bracket of $\Lambda$ with itself vanishes, that is, $\Lambda$ is a Poisson bivector field.
Consequently, we define a Poisson bracket among arbitrary smooth functions on $\mathfrak{T}_{1}(\Hh)$ by setting:

\be
\{F,\,G\}:=\Lambda(\mathrm{d}F,\,\mathrm{d}G)\,.
\ee
Then, we may define Hamiltonian vector fields associated with smooth functions on $\mathfrak{T}_{1}(\Hh)$.
Specifically, the Hamiltonian vector field associated with $F$ is the unique vector field $X_{F}$ such that (see \cite{marsden_ratiu-introduction_to_mechanics_and_symmetry} chapter $10$):

\be
X_{F}(G)=\,\{G,\,F\}=:\Lambda(\mathrm{d}G,\,\mathrm{d}F)\,.
\ee
Of particular importance are the Hamiltonian vector fields associated with the $f_{\mathbf{A}}$'s introduced above.
We denote these vector fields as $L^{\mathbf{A}}$, where $\mathbf{A}=\imath\mathbf{a}$ and $\Tr(\mathbf{a})=0$.
In this  case, we have:

\be\label{eqn: action of hamiltonian vector fields associated with linear functions on linear functions}
L^{\mathbf{A}}f_{\mathbf{B}}=\Lambda(\mathrm{d}f_{\mathbf{B}},\,\mathrm{d}f_{\mathbf{A}})=-f_{[\mathbf{A},\,\mathbf{B}]}\,.
\ee
Using the Jacobi identity for $\{\cdot,\cdot\}$ and the fact that Hamiltonian vector fields are derivation of $\{\cdot,\cdot\}$ it is possible to prove that (see \cite{abraham_marsden_ratiu-manifolds_tensor_analysis_and_applications} page $333$, \cite{marsden_ratiu-introduction_to_mechanics_and_symmetry} chapter $10$):

\be\label{eqn: fundamental vector fields of left action are an anti-representation}
[L^{\mathbf{A}},\,L^{\mathbf{B}}]=-L^{[\mathbf{A},\,\mathbf{B}]}\,,
\ee
which means that these vector fields close on an anti-realization of the Lie algebra $\suh$.
By changing sign in \ref{eqn: expectation value functions} we would obtain a realization instead of an antirealization.
These vector fields are complete because they are linear, and thus the anti-realization of $\suh$ ``integrates'' to the left action $\widetilde{\alpha}$ of $\SUh$ on $\mathfrak{T}_{1}(\Hh)$.
An explicit computation shows that the action $\widetilde{\alpha}$ is precisely that given in equation \eqref{eqn: action of SUh on quantum states}:

\be
(\mathbf{U},\,\xi)\mapsto \widetilde{\alpha}_{\mathbf{U}}(\xi)=\mathbf{U}\,\xi\,\mathbf{U}^{\dagger}\,.
\ee
Note again that this action preserves the convex structure of the space of quantum states $\stsp$.
Specifically, denoting with $\sum_{j}\,p_{j}\,\rho_{j}$ the convex combination of the quantum states $\rho_{j}$, we have:

\be
\widetilde{\alpha}_{\mathbf{U}}\left(\sum_{j}\,p_{j}\,\rho_{j}\right)=\sum_{j}\,p_{j}\,\widetilde{\alpha}_{\mathbf{U}}(\rho_{j})= \sum_{j}\,p_{j}\,\mathbf{U}\,\rho_{j}\,\mathbf{U}^{\dagger}\,.
\ee

According to the theory of actions of Lie groups we can introduce the notion of \grit{fundamental vector field} for the action of $\SUh$ on $\mathfrak{T}_{1}(\Hh)$:

\begin{definition}[\grit{Fundamental vector field} ]\label{defn: fundamental vector field}
Let $G$ be a Lie group, and let $\widetilde{\alpha}\colon G\times M\rightarrow M$ be a smooth left action of $G$ on the differential manifold $M$.
Given $\mathbf{A}\in\mathfrak{g}$, where $\mathfrak{g}$ is the Lie algebra of $G$, we define the fundamental vector field $X^{\mathbf{A}}\in\mathfrak{X}(M)$ as the infinitesimal generator of the flow $\widetilde{\alpha}^{\mathbf{A}}\colon \mathbb{R}\times M\rightarrow M$ given by

\be
\widetilde{\alpha}^{\mathbf{A}}(t\,,m):=\widetilde{\alpha}(\exp(t\mathbf{A})\,,m)\,,
\ee
that is:

\be
\left(X^{\mathbf{A}}f\right)(m)=\frac{d}{dt}\left.\,f\left(\widetilde{\alpha}_{\exp(t\mathbf{A})}(m)\right)\right|_{t=0}
\ee
for every smooth function $f$ on $M$.
\end{definition}

Among other things, the fundamental vector fields are useful in order to characterize the tangent space of the orbits of $\widetilde{\alpha}$ according to the following proposition  (see, for example, \cite{abraham_marsden_ratiu-manifolds_tensor_analysis_and_applications} page $331$):

\begin{proposition}\label{prop: characterization of tangent vectors on orbits by means of fundamental vector fields}
Let $G$ be a Lie group, and let $M$ be a smooth homogeneous space for $G$ according to the smooth action $\widetilde{\alpha}\colon G\times M\rightarrow M$.
Then, the tangent space $T_{m}M$ is:

\be
T_{m}M=\left\{X^{\mathbf{A}}(m)\,,\mathbf{A}\in\mathfrak{g}\right\}\,,
\ee
where $X^{\mathbf{A}}$ is the fundamental vector field defined in \ref{defn: fundamental vector field}.
\end{proposition}

Even though it is quite elementary, we want to show that $\widetilde{\alpha}$ is the action associated with the anti-realization of $\suh$ given by the Hamiltonian vector fields $L^{\mathbf{A}}$.
At this purpose, we compute the action of the fundamental vector field $X^{\mathbf{A}}$ of $\widetilde{\alpha}$ (see defintion \ref{defn: fundamental vector field}) on a generic linear function $f_{\mathbf{B}}$:

\be\label{eqn: fundamental vector fields of the unitary action on the affine hyperplane}
\begin{split}
X^{\mathbf{A}}f_{\mathbf{B}}(\xi)&=\frac{d}{dt}\left. f_{\mathbf{B}}\left(\widetilde{\alpha}_{\exp(t\mathbf{A})}(\xi)\right)\right|_{t=0}=\\
&=\frac{d}{dt}\left. f_{\mathbf{B}}\left(\mathrm{e}^{t\,\mathbf{A}}\,\xi\,\mathrm{e}^{-t\,\mathbf{A}}\right)\right|_{t=0}=\\
&=f_{\mathbf{B}}\left([\mathbf{A},\,\xi]\right)=-\imath\Tr\left(\mathbf{B}\,[\mathbf{A},\,\xi]\right)=\\
&=-\imath\Tr\left([\mathbf{B},\,\mathbf{A}]\,\xi\right)=-f_{[\mathbf{A},\,\mathbf{B}]}(\xi)\,.
\end{split}
\ee
Recalling equation \eqref{eqn: action of hamiltonian vector fields associated with linear functions on linear functions} we have:

\be
X^{\mathbf{A}}f_{\mathbf{B}}=L^{\mathbf{A}}f_{\mathbf{B}}\,,
\ee
and thus, since the differentials of the functions $f_{\mathbf{A}}$ with $\mathbf{A}\in\suh$ span the cotangent space at each point of $\mathfrak{T}_{1}(\Hh)$, we obtain $X^{\mathbf{A}}=L^{\mathbf{A}}$ from which we can conclude that the  vector fields $L^{\mathbf{A}}$ (with $\mathbf{A}\in\suh$) are the fundamental vector fields of the left action $\widetilde{\alpha}$ as claimed.

\begin{remark}[Coordinate expression]\label{rem: coordinate expression}
In the coordinate system introduced in remark \ref{rem: lower bound on purity depends on rank}, the Poisson bivector field $\Lambda$ reads:

\be\label{eqn: coordinate expression of poisson bivector}
\Lambda=\,c^{jk}_{l}\,x^{l}\,\frac{\partial}{\partial x^{j}}\,\wedge\,\frac{\partial}{\partial x^{k}}\,,
\ee
and the fundamental vector field  $L^{j}$ associated with  $\tau^{j}$ reads:

\be
L^{j} = c^{kj}_{l}\,x^{l}\,\frac{\partial}{\partial x^{k}}\,. 
\ee
If $f=a_{j}\,x^{j}$ is a linear function on the vector space $\suh$, the vector field will be $c^{jk}_{l}\,a_{j}x^{l}\,\frac{\partial}{\partial x^{k}}$.
\end{remark}

Because of the closed subgroup theorem (see \cite{abraham_marsden-foundations_of_mechanics} page $264$), the orbits of the action $\widetilde{\alpha}$ of $\SUh$ on $\mathfrak{T}_{1}(\Hh)$  are compact embedded submanifolds of $\mathfrak{T}_{1}(\Hh)$  labelled by the eigenvalues of the Hermitean operators that belong to them.
The tangent space $T_{\xi}\mathcal{O}$ at a point $\xi$ on the orbit $\mathcal{O}$ is characterized, according to proposition \ref{prop: characterization of tangent vectors on orbits by means of fundamental vector fields}, as the vector space spanned by the traceless Hermitean operators of the form $\imath[\mathbf{a},\,\xi]$ for some Hermitean $\mathbf{a}$.
Furthermore, every orbit $\mathcal{O}$ is a smooth homogeneous space for $\SUh$, and we write $\alpha\colon \SUh\times\mathcal{O}\rightarrow\mathcal{O}$:

\be
(\mathbf{U},\,\xi)\;\mapsto\;\alpha_{\mathbf{U}}(\xi):=\mathbf{U}\,\xi\,\mathbf{U}^{\dagger}
\ee
to denote the natural transitive left action of $\SUh$ on $\mathcal{O}$ (the ``restriction of $\widetilde{\alpha}$ to $\mathcal{O}$'').
Then, taking $\mathbf{A}=\imath\mathbf{a}$ in the Lie algebra $\suh$ of $\SUh$, the fundamental vector field for $\alpha$ associated with $\mathbf{A}$ is the ``restriction of $L^{\mathbf{A}}$ to $\mathcal{O}$'', and we will denote it as $X^{\mathbf{A}}$.
In the following, we will often use the representation of the tangent vector $T_{\xi}\mathcal{O}$ as a traceless Hermitean operator $\imath[\mathbf{a},\,\xi]$.
In this case, given an arbitrary vector field $X\in\mathfrak{X}(\mathcal{O})$, there will be a traceless Hermitean operator $\mathbf{a}$ such that the value of the vector field $X$ at the point $\xi$ may be written as $X(\xi)=\imath[\mathbf{a},\,\xi]$.
If $X\equiv X^{\mathbf{A}}$ is the fundamental vector field of $\alpha$ associated with $\mathbf{a}$, then  we have $X^{\mathbf{A}}(\xi)=\imath[\mathbf{a},\,\xi]$.

\begin{remark}[Manifolds of isospectral quantum states]
If we focus our attention on the orbits lying in the space of quantum states $\stsph$, that is, if we consider the orbits passing through $\rho\in\mathfrak{T}_{1}(\Hh)$ with $\rho$ positive-semidefinite, we have that the $n$-uple of eigenvalues of $\rho$ and $\mathbf{U}\,\rho\,\mathbf{U}^{\dagger}$ may be thought of as elements in the $n$-dimensional simplex $\Delta_{n}$, that is, the set of $n$-uple $(p_{1},...,p_{n})$ of non-negative real numbers such that $\sum_{j=1}^{n}\,p_{j}=1$.
On $\Delta_{n}$, the permutation group acts naturally, and we  denote with $\sigma$ the equivalence class $[\vec{p}]$ of $\vec{p}\in\Delta_{n}$ with respect to this action, and write $\Pi_{n}$ for the set of all these equivalence classes.
It is then clear that each orbit of $\SUh$ which is internal to the space of quantum states $\stsph$ can be labelled by a point $\sigma\in\Pi_{n}$ and we will denote it by:

\be
\stsp_{\sigma}(\Hh):=\left\{\rho\in\stsph\subset\mathfrak{T}_{1}(\Hh)\colon\;\left[\mathrm{sp}(\rho)\right]=\sigma\right\}\,,
\ee
where $\mathrm{sp}(\rho)$ is a probability vector the elements of which are the points in the spectrum of $\rho$.
The submanifolds $\stsp_{\sigma}(\Hh)$ are referred to as the submanifolds of isospectral quantum states.
It is customary  to represent an equivalence class by ordering the  elements of  the n-uple $(p_{1},...,p_{n})$ either in increasing or decreasing order.
\end{remark}

From the explicit expression \eqref{eqn: coordinate expression of poisson bivector} of $\Lambda$ we realize that the characteristic distribution of $\Lambda$, that is, the set of vector fields that are Hamiltonian with respect to $\Lambda$, is spanned by the generators $L^{\mathbf{A}}$ of the action of $\SUh$:

\be
X_{F}=i_{\mathrm{d}F}\Lambda= c^{jk}_{l}\,x^{l}\,\frac{\partial F}{\partial x^{k}}\,\frac{\partial }{\partial x^{j}}= \frac{\partial F}{\partial x^{k}}\,L^{k}\,.
\ee
Consequently, we conclude that the orbits of $\SUh$ on $\mathfrak{T}_{1}(\Hh)$ are precisely the symplectic leaves of the symplectic foliation associated with $\Lambda$ \cite{weinstein-the_local_structure_of_poisson_manifolds, weinstein-the_local_structure_of_poisson_manifolds_errata_and_addenda}.

\vsp

In addition to the action of the special unitary group $\SUh$, it has recently been shown that there is an action of the special linear group $\SLh$ on the space $\stsph$ of quantum states\footnote{As shown in \cite{ciaglia_dicosmo_ibort_laudato_marmo-dynamical_vector_fields_on_the_manifold_of_quantum_states} (remark $1$), this action is defined only on $\stsph$ and not on the whole affine hyperplane $\mathfrak{T}_{1}$.}  \cite{ciaglia_dicosmo_ibort_laudato_marmo-dynamical_vector_fields_on_the_manifold_of_quantum_states, ciaglia_ibort_marmo-differential_geometry_of_quantum_states_observables_and_evolution, grabowski_kus_marmo-geometry_of_quantum_systems_density_states_and_entanglement, grabowski_kus_marmo-symmetries_group_actions_and_entanglement, grabowski_kus_marmo-on_the_relation_between_states_and_maps_in_infinite_dimensions}. This action does not preserve the convex structure of $\stsph$, and it allows us to partition it into the disjoint union of smooth manifolds labelled by the rank of the quantum states (density operators) they contain.
Specifically, given an element $\gr$ in the special linear group $\SLh$, its action on a quantum state $\rho$ is given by:

\be\label{eqn: action of SLh on quantum states}
\rho\;\;\mapsto\;\; \frac{\gr\,\rho\,\gr^{\dagger}}{\Tr\left(\gr\,\rho\,\gr^{\dagger}\right)}\,.
\ee
The fact that this action does not preserve convex combinations of quantum states is to be ascribed to the presence of the trace term in the denominator.
Recalling that the special unitary group $\SUh$ is a (maximally compact) subgroup of $\SLh$, we see that the trace term becomes $1$ when $\gr$ is in $\SUh$ and we obtain again the action of the special unitary group $\SUh$ given by equation \eqref{eqn: action of SUh on quantum states}.

\begin{remark}
The action of $\SLh$ on $\stsph$ given by equation \eqref{eqn: action of SLh on quantum states} is not the only action of $\SLh$ on quantum states.
Indeed, in the following sections we will prove that there is a transitive action of $\SLh$ on every manifold of isospectral quantum states which is clearly different from the one given by equation \eqref{eqn: action of SLh on quantum states} unless we are considering pure quantum states.
\end{remark}

With a little effort   \cite{ciaglia_dicosmo_ibort_laudato_marmo-dynamical_vector_fields_on_the_manifold_of_quantum_states, ciaglia_ibort_marmo-differential_geometry_of_quantum_states_observables_and_evolution, grabowski_kus_marmo-geometry_of_quantum_systems_density_states_and_entanglement, grabowski_kus_marmo-symmetries_group_actions_and_entanglement, grabowski_kus_marmo-on_the_relation_between_states_and_maps_in_infinite_dimensions} it is possible to prove that the orbits of $\SLh$ on $\stsph$ by means of the action given in equation \eqref{eqn: action of SLh on quantum states} are labelled by the rank of the quantum states they contain, and that they are differential manifolds.
We will denote these orbits as $\stsp_{k}(\Hh)$, where $k$ denotes the rank of the quantum states in $\stsp_{k}(\Hh)$.
In particular, the manifold $\stsp_{1}(\Hh)$ coincides with the space of pure quantum states.
This means that the pure quantum states form a homogeneous space for both the special unitary group $\SUh$ and for its complexificiation, the complex special linear group $\SLh$.

If we select an orthonormal basis in $\Hh$, then  we may take as a representative sate for every orbit $\stsp_{k}(\Hh)$ the following quantum state:

\be
\rho_{k}=\frac{1}{k}\,\left(\begin{matrix} \mathbb{I}_{k} & \mathbf{0} \\ \mathbf{0} & \mathbf{0} \\\end{matrix}\right)\,,
\ee
where $\mathbb{I}_{k}$ is the identity $(k\times k)$ matrix with respect to the chosen orthonormal basis in $\Hh$.

For each k, the orbit $\stsp_{k}(\Hh)$ will be a face of the convex body of quantum states, and we find that the topological boundary will not be a smooth manifold because it is the union of submanifolds of changing dimension \cite{grabowski_kus_marmo-symmetries_group_actions_and_entanglement}.

\vsp

According to \cite{ciaglia_dicosmo_ibort_laudato_marmo-dynamical_vector_fields_on_the_manifold_of_quantum_states} we can give an infinitesimal description of the action of $\SLh$ on $\stsph$ in terms of vector fields on the whole $\mathfrak{T}_{1}(\Hh)$.
These vector fields are divided into Hamiltonian and gradient-like vector fields.
The Hamiltonian vector fields are precisely the vector fields $L^{\mathbf{A}}$ with $\mathbf{A}\in\suh$ generating the action of $\SUh$ and, as we have previously seen, are defined by means of an anti-symmetric bivector field on $\mathfrak{T}_{1}(\Hh)$, the Poisson tensor $\Lambda$  coming from the Lie algebra structure of $\suh$.
Regarding the gradient-like vector fields $\widetilde{Y}^{\mathbf{A}}$ with $\mathbf{A}\in\suh$, as anticipated above, they can be defined starting from a symmetric contravariant tensor field $\mathcal{R}$   on $\mathfrak{T}_{1}(\Hh)$.
Recalling once again that the exterior differentials of the functions $f_{\mathbf{A}}$ with $\mathbf{A}\in\suh$ generate the cotangent space at each point of $\mathfrak{T}_{1}(\Hh)$, the tensor field $\mathcal{R}$ is defined by:

\be\label{eqn: symmetric bivector preserving the rank}
\mathcal{R}\left(\mathrm{d}f_{\mathbf{A}}\,,\mathrm{d}f_{\mathbf{B}}\right) = f_{\mathbf{A}\odot\mathbf{B}} - 2f_{\mathbf{A}}\,f_{\mathbf{B}}\,,
\ee
where:

\be
\mathbf{A}\odot\mathbf{B} = - \imath\,\left\{\mathbf{A},\,\mathbf{B}\right\}= - 	\imath\,(\mathbf{A}\,\mathbf{B} + \mathbf{B}\,\mathbf{A}) 
\ee
is the symmetric product coming from the anti-commutator in $\Bh$, and then extended by linearity.

It should be noticed that the right hand side of \eqref{eqn: symmetric bivector preserving the rank} gives the variance when $\mathbf{A}=\mathbf{B}$ and the covariance in the general case.
The expression of $\mathcal{R}$ in the coordinate system $\{x^{j}\}_{j=1,...,n^{2}-1}$ defined in remark \ref{rem: lower bound on purity depends on rank} reads:

\be\label{eqn: R on trace 1}
\mathcal{R}=\left(d^{jk}_{l}\,x^{l} + d_{0}^{jk}\right)\,\frac{\partial}{\partial x^{j}}\otimes\frac{\partial}{\partial x^{k}} -   2\Delta\otimes\Delta\,,
\ee  
where the coefficients $d^{jk}_{l}$ and $d^{jk}_{0}$ are defined according to:

\be
\tau^{j}\odot\tau^{k}= d^{jk}_{l}\tau^{l} + d^{jk}_{0}\mathbb{I}\,.
\ee
Analogously to what happens for the Hamiltonian vector fields, the gradient-like vector field $\widetilde{Y}_{F}$ associated with the smooth function $F$ is defined as the vector field such that:

\be
\widetilde{Y}_{F}(G)=\mathcal{R}(\mathrm{d}F,\,\mathrm{d}G)
\ee
for every smooth function $G$ on $\mathfrak{T}_{1}(\Hh)$.
In particular, the gradient-like vector field $\widetilde{Y}_{f_{\mathbf{A}}}\equiv\widetilde{Y}^{\mathbf{A}}$ associated with $f_{\mathbf{A}}$ reads:

\be\label{eqn: gradient-like vector field on trace 1}
\widetilde{Y}^{\mathbf{A}} = \left(d^{jk}_{l}\,x^{l}a_{j} +   \frac{\delta^{jk}a_{j}}{n} \right)\,\frac{\partial}{\partial x^{k}} -  2x^{j}a_{j} \,\Delta\,,
\ee
where $\suh\ni\mathbf{A}=a_{j}\,\tau^{j}$.
In general, these gradient-like vector fields are not complete on $\mathfrak{T}_{1}(\Hh)$.
However, as it is shown in \cite{ciaglia_dicosmo_ibort_laudato_marmo-dynamical_vector_fields_on_the_manifold_of_quantum_states},  the integral curve $\gamma_{\rho}(t)$ of the vector field $L^{\mathbf{A}} + \widetilde{Y}^{\mathbf{B}}$ starting at any $\rho\in\stsph$ reproduces the non-convex action of $\SLh$ on $\stsp$ given by equation \eqref{eqn: action of SLh on quantum states}, that is, $\gamma_{\rho}(t)$ reads:

\be
\gamma_{\rho}(t)=\frac{\gr_{t}\,\rho\,\gr^{\dagger}_{t}}{\Tr\left(\gr_{t}\,\rho\,\gr^{\dagger}_{t}\right)}
\ee
with $\gr_{t}=\mathrm{e}^{t(\mathbf{B} - \imath\mathbf{A})}$.

\vsp

Up to now, we considered only contravariant bivector fields on $\mathfrak{T}_{1}(\Hh)$ ``coming from'' the anti-symmetric and the symmetric product in $\mathcal{B}(\Hh)$.
However, the $C^{*}$-algebra $\mathcal{B}(\Hh)$ also carries the structure of a Hilbert space with the Hilbert-Schimdt product between $\mathbb{A}\,,\mathbb{B}$ given by:

\be
\langle\mathbb{A}|\mathbb{B}\rangle = \Tr\left(\mathbb{A}^{\dagger}\,\mathbb{B}\right)\,,
\ee
and, since the affine hyperplane $\mathfrak{T}_{1}(\Hh)$ is a subset in $\mathcal{B}(\Hh)$, we may pullback the Hermitean tensor associated with the Hilbert product $\langle\cdot|\cdot\rangle$ on $\mathcal{B}(\mathcal{H})$ along the canonical immersion of $\mathfrak{T}_{1}(\Hh)$ in $\mathcal{B}(\Hh)$.
The symmetric (real) part of the resulting tensor field gives rise to a Riemannian metric  $\mathcal{N}$ on $\mathfrak{T}_{1}(\Hh)$.
In the coordinate system introduced above in remark \ref{rem: lower bound on purity depends on rank}, this metric tensor field would be:

\be\label{eqn: Riemannian metric tensor on affine hyperplane}
\mathcal{N}=\delta_{jk}\,\mathrm{d}x^{j}\,\otimes\,\mathrm{d}x^{k}\,.
\ee
We notice that $\mathcal{N}$ is invariant under the action of Hamiltonian vector fields.
This follows  directly from the Adjoint-invariance of the Hilbert-Schmidt  product.
Clearly, we may pullback the metric tensor $\mathcal{N}$ to an orbit $\mathcal{O}$ of $\SUh$ or $\SLh$ obtaining a metric tensor on the orbits which is invariant with respect to the action of $\SUh$.
In the case of the space of pure quantum states, the invariance under $\SUh$ implies that the resulting metric tensor must be proportional to the Fubini-Study metric tensor (see chapter $4$ in \cite{bengtsson_zyczkowski-geometry_of_quantum_states:_an_introduction_to_quantum_entanglement}, and \cite{ercolessi_marmo_morandi-from_the_equations_of_motion_to_the_canonical_commutation_relations}).

\section{K\"{a}hler structures on isospectral orbits}\label{sec: Kahler structures on the manifolds of isospectral quantum states}

All the orbits of $\SUh$ in $\mathfrak{T}_{1}(\Hh)$, exception made for the degenerate orbit passing through the maximally mixed state $\rho_{m}=\frac{\mathbb{I}}{n}$, possess a very rich geometrical structure: they are K\"ahler manifolds   \cite{ercolessi_marmo_morandi-from_the_equations_of_motion_to_the_canonical_commutation_relations, grabowski_kus_marmo-geometry_of_quantum_systems_density_states_and_entanglement}. This means that, on each isospectral orbit $\mathcal{O}$, there is a symplectic form $\omega$, a Riemannian metric tensor $g$ and a complex structure\footnote{A complex structure $J$ on a manifold $M$ is a $(1,1)$ tensor field such that $J\circ J=-\mathrm{Id}$, and such that its Nijenhuis tensor vanishes.
For more details, we refer to \cite{bengtsson_zyczkowski-geometry_of_quantum_states:_an_introduction_to_quantum_entanglement, marmo_ferrario_lovecchio_morandi_rubano-the_inverse_problem_in_the_calculus_of_variations_and_the_geometry_of_the_tangent_bundle, nirenberg_newlander-complex_analytic_coordinates_in_almost_complex_manifolds}.} $J$ satisfying a particular compatibility condition.
Furthermore, all these three tensor fields are invariant with respect to the action of $\SUh$ generating the orbit.
The Riemannian metric tensor $g$ ``comes'' from the Killing form of $\SUh$ (see remark \ref{rem: isospectral orbits as homogeneous spaces}), and thus, since $\SUh$ is a simple Lie group, it follows that every Riemannian metric tensor on the isospectral orbit $\mathcal{O}$ which is invariant with respect to the action of $\SUh$ must be a constant multiple of $g$.

The following theorem can be proved to hold (see \cite{grabowski_kus_marmo-geometry_of_quantum_systems_density_states_and_entanglement} theorem $7$, theorem $8$ and the paragraph just after theorem $8$):

\begin{theorem}\label{thm: isospectral quantum states as kahler manifolds}
Let $\mathcal{O}\subset\mathfrak{T}_{1}(\Hh)$ be an orbit of $\SUh$ not passing through the maximally mixed state $\rho_{m}=\frac{\mathbb{I}}{n}$.
Then, $\mathcal{O}$ is a K\"{a}hler manifold.
This means that there are a symplectic form $\omega$, a Riemannian metric tensor $g$ and a complex structure $J$ such that:

\be
g\left(X\,,Y\right)=\omega\left(X\,,J(Y)\right)\,,\;\;\;\;\forall\;\;X,Y\in\mathfrak{X}(\mathcal{O}).
\ee
All of these tensors are invariant with respect to the canonical action of $\SUh$ on the orbit, that is:

\be
\phi_{\mathbf{U}}^{*}\omega =\omega\,,\;\;\; \phi_{\mathbf{U}}^{*}g =g\,,\;\;\; \phi_{\mathbf{U}}^{*}J =J\,,
\ee
where $\phi_{\mathbf{U}}$ is the diffeomorphism of $\mathcal{O}$ associated with the action of $\mathbf{U}\in \SUh$.
From the infinitesimal point of view, denoting by $X^{\mathbf{A}}$ the generic fundamental vector field of the action of $\SUh$ on $\mathcal{O}$, we have:

\be
\mathcal{L}_{X^{\mathbf{A}}}\,(\omega )=0\,,\;\;\;\mathcal{L}_{X^{\mathbf{A}}}\,(g )=0\,,\;\;\;\mathcal{L}_{X^{\mathbf{A}}}\,(J )=0\,,\;\;\;
\ee
where $\mathcal{L}$ denotes the Lie derivative.

The symplectic form on every $\mathcal{O}$ is given by:

\be\label{eqn: symplectic structure on isospectral states}
(\omega (X\,,Y))(\xi)=\omega_{\xi}\left(\imath[\mathbf{a}\,,\xi]\,,\imath[\mathbf{b}\,,\xi]\right):=\imath\,\Tr\left([\mathbf{a}\,,\xi]\,\mathbf{b}\right)=\imath\,\Tr\left(\xi\,[\mathbf{b}\,,\mathbf{a}]\right)\,,
\ee
where $X(\xi)=\imath[\mathbf{a}\,,\xi\,]$ and $Y(\xi)=\imath[\mathbf{b}\,,\xi\,]$ are tangent vectors at $\xi$.

Let us denote with $\{|j\rangle\}_{j=1,...,n}$ the basis of eigenvectors of $\xi\in\mathcal{O}$, and let us order the basis elements so that $\lambda^{1}\geq\lambda^{2}\geq\cdots\geq\lambda^{n}$, where $\lambda^{j}$ is the $j$-th eigenvalue of $\xi$.
Define the Weyl basis $\mathbf{M}_{kl}:=|k\rangle\langle l|$ with $k<l$, and $\mathbf{D}_{j}:=|j\rangle\langle j|$.
Then, the complex structure on every $\mathcal{O}$ is given by:

\be
\left(J(X)\right)(\xi)=J_{\xi}(\imath[\mathbf{a}\,,\xi]):=\sum_{k<l}\,\left(\lambda^{k} - \lambda^{l}\right)\,\left(a^{kl}\mathbf{M}_{kl} + \bar{a}^{kl}\mathbf{M}_{lk}\right)\,,
\ee
where $\mathbf{a}=a^{kl}\mathbf{M}_{kl} + \bar{a}^{kl}\mathbf{M}_{lk} + a^{j}\mathbf{D}_{j}$.
Accordingly, the symplectic form $\omega$ and the metric tensor $g$ on every $\mathcal{O}$ are given by:

\be
(\omega(X\,,Y))(\xi) = \omega_{\xi}\left(\imath[\mathbf{a}\,,\xi]\,,\imath[\mathbf{b}\,,\xi]\right)=\sum_{k<l}\,\imath\,\left(\lambda^{l} - \lambda^{k}\right)\,\left(a^{kl} \bar{b}^{kl} - \bar{a}^{kl}b^{kl}\right)\,,
\ee
\be
\begin{split}
(g(X\,,Y))(\xi)&=g_{\xi}\left(\imath[\mathbf{a}\,,\xi]\,,\imath[\mathbf{b}\,,\xi]\right)= \omega_{\xi}\left(\imath[\mathbf{a}\,,\xi]\,,J_{\xi}\left(\imath[\mathbf{b}\,,\xi]\right)\right)=\\
&=\sum_{k<l}\,\left(\lambda^{k} - \lambda^{l}\right)\,\left(a^{kl} \bar{b}^{kl} + \bar{a}^{kl}b^{kl}\right)\,,
\end{split}
\ee
where   $\mathbf{a}=a^{kl}\mathbf{M}_{kl} + \bar{a}^{kl}\mathbf{M}_{lk} + a^{j}\mathbf{D}_{j}$, $Y(\rho)=\imath[\mathbf{b}\,,\bar{\rho}]$, and $\mathbf{b}=b^{kl}\mathbf{M}_{kl} + \bar{b}^{kl}\mathbf{M}_{lk} + b^{j}\mathbf{D}_{j}$.
\end{theorem}

\begin{remark}\label{rem: complex structure on pure quantum states}
An explicit computation shows that, when we consider the orbit corresponding to the space $\stsp_{1}(\Hh)$ of pure quantum states, that is, the rank-one projectors in $\mathfrak{T}_{1}(\Hh)$, the action of the complex structure $J$ on a vector field $X$ evaluated at the pure quantum state $\rho$ reads:

\be
\left(J(X)\right)(\rho)=[[\mathbf{a}\,,\rho]\,,\rho]=\{\mathbf{a}\,,\rho\} - 2 \rho\,\mathbf{a}\,\rho\,,
\ee
where $X(\rho)=\imath[\mathbf{a},\,\rho]$, and $\{\cdot\,,\cdot\}$ is the matrix anticommutator.
Consequently, the action of the metric tensor $g$ on the vector fields $X$ and $Y$ evaluated at $\rho$ reads:

\be
\left(g\left(X\,,Y\right)\right)(\rho)=\Tr\left([\mathbf{a}\,,\rho]\,[\mathbf{b}\,,\rho]\right)\,
\ee
Since we are on the space of pure quantum states which is diffeomorphic to the complex projective space $\mathbb{CP}(\Hh)$, and since $g$ is clearly invariant under the action of $\SUh$, it follows that $g$ must be proportional to the Fubini-Study metric tensor (see chapter $4$ in \cite{bengtsson_zyczkowski-geometry_of_quantum_states:_an_introduction_to_quantum_entanglement}, and \cite{ercolessi_marmo_morandi-from_the_equations_of_motion_to_the_canonical_commutation_relations}).
\end{remark}

\begin{remark}\label{rem: isospectral orbits as homogeneous spaces}
Since every orbit $\mathcal{O}$ is a homogeneous space of the special unitary group $\SUh$, we may always find a diffeomorphism $\mathcal{O}\cong \SUh/G_{\rho}$, where $G_{\rho}\subset\SUh$ is the isotropy subgroup of some fiducial $\rho\in\mathcal{O}$.
In this way, we may consider the canonical projection map:

\be
\pi_{\rho}\colon\,\SUh\,\rightarrow\,\SUh/G_{\rho}\cong\mathcal{O}\,,
\ee
and we obtain the pullback $\pi_{\rho}^{*}g$ of the invariant metric tensor $g$ on $\SUh$ \cite{aniello_clemente-gallardo_marmo_volkert-classical_tensors_and_quantum_entanglement_I}:

\be
\pi_{\rho}^{*}g=\Tr\left(\rho\,\gr^{-1}\,\mathrm{d}\gr\otimes \gr^{-1}\,\mathrm{d}\gr\right)\,,
\ee
where $\gr^{-1}\,\mathrm{d}\gr$ is the Maurer-Cartan form of $\SUh$.
Analogously, the pullback $\pi_{\rho}^{*}\omega$ of the symplectic form $\omega$ reads:

\be
\pi_{\rho}^{*}\omega=\Tr\left(\rho\,\gr^{-1}\,\mathrm{d}\gr\wedge \gr^{-1}\,\mathrm{d}\gr\right)\,,
\ee
which turns out to be an exact one-form unlike the one we started from on the orbit \cite{ercolessi_marmo_morandi_mukunda-geometry_of_mixed_states_and_degenerancy_structure_of_geometric_phases_for_multi-level_quantum_systems_a_unitary_group_approach}.
These structures turn out to be directly related to the structures arising when considering the Berry phase on the space of density matrices.
\end{remark}

Now that we have a K\"{a}hler structure on every (non-degenerate) orbit $\mathcal{O}$, we can define the Hamiltonian and gradient vector fields associated with smooth functions on $\mathcal{O}$.
Specifically, if $f$ is a smooth function on $\mathcal{O}$, we define its associated Hamiltonian vector field $X_{f}$ and its associated gradient vector field $Y_{f}$ to be, respectively, the vector fields given by:

\be
X_{f}:=\omega^{-1}\left(\mathrm{d}f\,,\cdot\right)\,,
\ee
\be
Y_{f}:=g^{-1}\left(\mathrm{d}f\,,\cdot\right)\,,
\ee
where $\omega$ and $g$ are, respectively, the canonical symplectic structure and canonical metric tensor of the K\"{a}hler structure on $\mathcal{O}$.
Note that, because of the compatibility condition between $\omega,g$, and $J$ characteristic of K\"{a}hler manifolds, we may write every gradient vector field $Y_{f}$ as:

\be
Y_{f}=J(X_{f})\,,
\ee
where $X_{f}$ is the Hamiltonian vector field associated with $f$ by means of the symplectic structure $\omega$.

Quite naturally, the fundamental vector fields associated with the action of $\SUh$ on $\mathcal{O}$ are the Hamiltonian vector fields associated with the pullback of the linear functions on $\mathfrak{T}_{1}(\Hh)$ by means of the canonical immersion $\mathrm{i}_{\mathcal{O}}\colon \mathcal{O}\rightarrow \mathfrak{T}_{1}(\Hh)$: 

\be
e_{\mathbf{A}}(\xi):=\mathrm{i}_{\mathcal{O}}^{*}f_{\mathbf{A}}(\xi)=-\imath\Tr\left(\xi\,\mathbf{A}\right)\,.
\ee
Indeed, we can prove the following:

\begin{proposition}\label{prop: fundamental vector fields and expectation value functions on the manifold isospectral states}
Let $\mathbf{A}=\imath\mathbf{a}$ be in the Lie algebra of $\SUh$, and consider the fundamental vector field $X^{\mathbf{A}}$ on the orbit $\mathcal{O}$.
Then, for every expectation value function $e_{\mathbf{B}}$ associated with $\mathbf{B}$ as defined above, the following equality holds:

\be
\mathcal{L}_{X^{\mathbf{A}}}\,(e_{\mathbf{B}})=e_{\imath[\mathbf{B}\,,\mathbf{A}]}\,,
\ee
from which it follows that $X^{\mathbf{A}}$ is the Hamiltonian vector field associated with $-e_{\mathbf{A}}$ by means of the symplectic structure $\omega$, that is:

\be
\omega\left(X^{\mathbf{A}}\,,\cdot\right)=-\mathrm{d}e_{\mathbf{A}}\,.
\ee
Then, every cotangent vector in $T^{*}_{\xi}\mathcal{O}$ may be represented as $\mathrm{d}e_{\mathbf{A}}(\xi)$ for some $\mathbf{A}\in\suh$.

\begin{proof*}
By the very definition of Lie derivative of a function with respect to a vector field we have:

\be\label{eqn: hamiltonian vector field and expectation value functions on isospectral states}
\begin{split}
\left(\mathcal{L}_{X^{\mathbf{A}}}\,(e_{\mathbf{B}})\right)(\xi)&=\left(X^{\mathbf{A}}\,e_{\mathbf{B}}\right)(\xi)=\left(\frac{\mathrm{d}}{\mathrm{d}t}\left(e_{\mathbf{B}}\left(\alpha_{\exp(t\mathbf{A})}(\xi)\right)\right)\right)_{t=0}=\\
&=-\imath\,\left(\frac{\mathrm{d}}{\mathrm{d}t}\,\left(\Tr\left(\exp( t\mathbf{A})\,\xi\,\exp(- t\mathbf{A})\,\mathbf{B}\right)\right)\right)_{t=0}=\\
&=-\imath\, \Tr\left(\left[\mathbf{A}\,,\xi\right]\,\mathbf{B}\right)=-\imath\,\Tr\left(\xi\,\left[\mathbf{B}\,,\mathbf{A}\right]\right)=e_{[\mathbf{B}\,,\mathbf{A}]}(\xi)\,,
\end{split}
\ee
and the first assertion is proved.
The fact that $X^{\mathbf{A}}$ is the Hamiltonian vector field associated with $e_{\mathbf{A}}$ when $\mathbf{A}=\imath\mathbf{a}$ is in $\suh$, can be seen as follows.
Every tangent vector $v_{\xi}\in T_{\xi}\mathcal{O}$ can be represented as $\imath[\mathbf{b}\,,\xi]$ for some $\mathbf{B}=\imath\mathbf{b}$  in $\suh$.
Consequently, according to equations \eqref{eqn: hamiltonian vector field and expectation value functions on isospectral states} and \eqref{eqn: symplectic structure on isospectral states}, we have

\be
\begin{split}
-\omega_{\xi}(X^{\mathbf{A}}(\xi)\,,\mathbf{v}_{\xi})&=-\omega_{\xi}(X_{\mathbf{A}}(\xi)\,,\imath[\mathbf{b},\,\xi])=\\
&=\imath\,\Tr\left(\xi\,\left[\mathbf{A}\,,\mathbf{B}\right]\right)=\\
&=-\left(X^{\mathbf{B}}\,e_{\mathbf{A}}\right)(\xi) =\\
&=-\left(\mathrm{d}e_{\mathbf{A}}(\xi)\right)(\mathbf{v}_{\xi})\,,
\end{split}
\ee
and since $\mathbf{v}_{\xi}$ is arbitrary, we conclude that:

\be
\omega\left(X^{\mathbf{A}}\,,\cdot\right)=-\mathrm{d}e_{\mathbf{A}}\,,
\ee
which means that $X^{\mathbf{A}}$ is the Hamiltonian vector field associated with the expectation value function $-e_{\mathbf{A}}$.
Clearly, the symplectic structure $\omega_{\xi}$ provides us with an isomorphism between the tangent space $T_{\xi}\mathcal{O}$ and the cotangent space $T^{*}_{\xi}\mathcal{O}$ for every $\xi\in\mathcal{O}$, therefore,  every cotangent vector in $T^{*}_{\xi}\mathcal{O}$ may be represented as $\mathrm{d}e_{\mathbf{A}}(\xi)$ for some $\mathbf{A}\in\suh$.
\end{proof*}
\end{proposition}

According to proposition \ref{prop: fundamental vector fields and expectation value functions on the manifold isospectral states}, we will write $X^{\mathbf{A}}$ and $Y^{\mathbf{A}}$ to denote, respectively, the Hamiltonian and gradient vector field associated with the expectation value function $-e_{\mathbf{A}}$.
The flow of Hamiltonian vector fields $X^{\mathbf{A}}$ may be dynamically interpreted as the unitary evolution of the quantum system.
These dynamical evolutions describe the behaviour of a closed quantum system, that is, a system which is isolated from its environment.
Consequently, from the point of view of the dynamics of closed quantum systems, Hamiltonian vector fields on the manifolds of isospectral states provide the correct geometrical framework for a complete treatment of the subject.
In the case of pure quantum states, this geometrical picture can be generalized to the infinite-dimensional case according to the work  \cite{cirelli_mania_pizzocchero-quantum_mechanics_as_an_infinite_dimensional_Hamiltonian_system_with_uncertainty_structure, cirelli_pizzocchero-on_the_integrability_of_quantum_mechanics_as_an_infinite_dimensional_system}.

The gradient vector fields, on the other hand, would represent nonlinear dissipative dynamical vector fields.
Their associated flows are transversal to the level sets of the expectation value function and would take from one eigenstate to a different eigenstate. 
In the q-bit case it would take from the ``north-pole'' to the ``south-pole''.

In the following, we will see that Hamiltonian and gradient vector fields associated with elements in the Lie algebra $\suh$ of $\SUh$ close on an anti-realization of the Lie algebra $\mathfrak{sl}(\mathcal{H})$ of the Lie group $\mathcal{SL}(\Hh)$.
On each orbit $\mathcal{O}$ of $\SUh$, this anti-realization of $\mathfrak{sl}(\mathcal{H})$ ``integrates'' to a left action of $\SLh$.

\begin{remark}[Lie algebroid on isospectral quantum states]
On every orbit $\mathcal{O}$, we may build a Lie-algebroid \cite{mackenzie-general_theory_of_lie_groupoids_and_algebroids} using the expectation value functions $e_{\mathbf{A}}$ and the Hamiltonian vector fields $X^{\mathbf{A}}$ as follows.
Consider the vector bundle $T^{*}\mathcal{O}$ over $\mathcal{O}$.
Take the sub-vector bundle

\be
\mathcal{E}:=\left\{\beta\in T^{*}\mathcal{O}\colon\;\beta\in\mathrm{Im}(\mathrm{d}e_{\mathbf{A}})\;\mbox{for some } \mathbf{A}\in\suh\right\}\,,
\ee
where $\mathrm{d}e_{\mathbf{A}}\colon\,\mathcal{O}\rightarrow T^{*}\mathcal{O}$ is looked at as a section of $T^{*}\mathcal{O}$.
Because of proposition \ref{prop: fundamental vector fields and expectation value functions on the manifold isospectral states}, the set of all the $\mathrm{d}e_{\mathbf{A}}$ is a basis for the module of the sections of the vector bundle $\mathcal{E}$.
Then, we define the anchor map $\tau\colon \mathcal{E}\rightarrow T\mathcal{O}$ as the linear (in the sense of modules) map such that:

\be
\tau(\mathrm{d}e_{\mathbf{A}}):=-\mathrm{i}_{\mathrm{d}e_{\mathbf{A}}}\,\omega^{-1} = X^{\mathbf{A}}\,.
\ee
Now, we define an antisymmetric product $[,]$ on the sections of $\mathcal{E}$ by setting:

\be
[\mathrm{d}e_{\mathbf{A}},\,\mathrm{d}e_{\mathbf{B}}]:=\mathrm{d}e_{[\mathbf{A},\,\mathbf{B}]}\,,
\ee
 on the basis elements (in the sense of modules) of the sections of $\mathcal{E}$.
denoting by $f$ a smooth function on $\mathcal{O}$, it is a matter of straightforward computation to show that $[,]$ satisfies the Jacobi identity and is such that:

\be
\begin{split}
[\Gamma_{1},\,f\Gamma_{2}]&=f\,[\Gamma_{1},\,\Gamma_{2}] + \tau(\Gamma_{1})(f)\,\Gamma_{2}\\
\tau\left([\Gamma_{1},\,\Gamma_{2}]\right)&=[\tau(\Gamma_{1}),\,\tau(\Gamma_{2})]\,, 
\end{split}
\ee
which means that $(\mathcal{E},\,[,],\,\tau)$ is a Lie-algebroid.
\end{remark}

\section{The q-bit I}\label{sec: q-bit I}

To help the reader gain some intuition on the structures presented so far, we will consider the explicit example of a two-level quantum system, i.e., the q-bit.
In this case, every Hermitean operator $\xi$ on $\Hh$ may be written as:

\be
\xi=\frac{1}{2}\left(\xi^{0}\,\sigma_{0} + \vec{\xi}\cdot\vec{\sigma}\right)\,,
\ee
where $\sigma_{0}=\mathbb{I}$ is the identity operator, and $\vec{\sigma}=(\sigma_{1},\,\sigma_{2},\,\sigma_{3})$ with $\sigma_{j}$ the $j$-th Pauli matrix.
Clearly, the affine hyperplane $\mathfrak{T}_{1}(\Hh)$ is identified by the constraint $\xi^{0}=1$, while it is well-known that the space of quantum states $\stsp$ is identified by the constraints:

\be
\rho^{0}=1,\;\;\;\;||\vec{\rho}||^{2}\leq 1\,,
\ee
that is, $\stsph$ is the so-called Bloch ball in $\mathfrak{T}_{1}(\Hh)$.

A basis in the Lie algebra $\suh$ of $\SUh$ is given by $\{\tau^{j}\}_{j=1,2,3}$ with $\tau^{j}=\imath\,\sigma_{j}$.
The structure constants are then given by:

\be
[\tau^{j},\,\tau^{k}]=-2\,\epsilon^{jk}_{l}\,\tau^{l}\,,
\ee
where $\epsilon^{jkl}=\delta^{lr}\,\epsilon^{jk}_{r}$ is the Levi-Civita symbol.
Accordingly, the coordinate system associated with $\{\tau^{j}\}_{j=1,2,3}$ is given by:

\be
x^{j}(\xi):=-\imath\,\Tr\left(\tau^{j}\,\xi\right)=\Tr\left(\sigma_{j}\,\xi\right)=\rho^{j}\,,
\ee
while the Poisson tensor $\Lambda$ reads:

\be
\Lambda=\,-2\,\epsilon^{jk}_{l}\,x^{l}\,\frac{\partial}{\partial x^{j}}\,\wedge\,\frac{\partial}{\partial x^{k}}\,,
\ee
and the fundamental vector fields  read:

\be
\begin{split}
L^{1} &= 2\,\left(x^{3}\,\frac{\partial}{\partial x^{2}} - x^{2}\,\frac{\partial}{\partial x^{3}}\right) \\
L^{2} &= 2\,\left(x^{1}\,\frac{\partial}{\partial x^{3}} - x^{3}\,\frac{\partial}{\partial x^{1}}\right) \\ 
L^{3} &= 2\,\left(x^{2}\,\frac{\partial}{\partial x^{1}} - x^{1}\,\frac{\partial}{\partial x^{2}}\right)\,. 
\end{split}
\ee
An explicit computation shows that:

\be
L^{j}(r^{2})=L^{j}(\delta_{kl}\,x^{k}\,x^{l})=0\,,
\ee
that is, the fundamental vector fields of $\SUh$ are tangent to the spheres in $\mathfrak{T}_{1}(\Hh)$ centered at $\frac{\mathbb{I}}{2}$.
Furthermore, exploiting the fact that the action of $\SUh$ preserves the eigenvalues, it can be proved that these spheres are precisely the orbits of $\SUh$ in $\mathfrak{T}_{1}(\Hh)$.
Specifically, the manifolds of isospectral quantum states are the spheres with $r\leq 1$, where the sphere with $r=1$ being the space of pure quantum states for the q-bit.

\vsp

Recalling what has been said at the end of section \ref{sec: Quantum states and the special unitary group}, we have an action of $\mathcal{SL}(\Hh)$ on the space $\stsp$ of quantum states of the q-bit.
This action does not preserve the spectrum of quantum states, but, it preserves the rank.
The infinitesimal description of this action is given in terms of vector fields on $\mathfrak{T}_{1}(\Hh)$.
These vector fields are build with the help of the Poisson bivector field $\Lambda$ introduced before, and with the help of the symmetric bivector field:

\be
\mathcal{R}=2\delta^{jk}\,\frac{\partial}{\partial x^{j}}\otimes\frac{\partial}{\partial x^{k}} -   2\Delta\otimes\Delta\,,
\ee 
as from equation \eqref{eqn: R on trace 1}.
In particular, by means of $\Lambda$ we obtain again the Hamiltonian vector fields:

\be
\begin{split}
L^{1} &= 2\,\left(x^{3}\,\frac{\partial}{\partial x^{2}} - x^{2}\,\frac{\partial}{\partial x^{3}}\right)\\
L^{2}&= 2\,\left(x^{1}\,\frac{\partial}{\partial x^{3}} - x^{3}\,\frac{\partial}{\partial x^{1}}\right) \\ 
L^{3} &= 2\,\left(x^{2}\,\frac{\partial}{\partial x^{1}} - x^{1}\,\frac{\partial}{\partial x^{2}}\right)
\end{split}
\ee
generating the action of $\SUh$ on $\mathfrak{T}_{1}(\Hh)$.
While, by means of the symmetric tensor $\mathcal{R}$, we obtain the gradient-like vector fields:

\be
\begin{split}
\widetilde{Y}^{1} & =2\,\left(x^{1}\,\Delta - \frac{\partial }{\partial x^{1}}\right) \\
\widetilde{Y}^{2} & =2\,\left(x^{2}\,\Delta - \frac{\partial }{\partial x^{2}}\right) \\
\widetilde{Y}^{3} & =2\,\left(x^{3}\,\Delta - \frac{\partial }{\partial x^{3}}\right)\,.
\end{split}
\ee
A direct computation shows that $(L^{1},L^{2},L^{3},\widetilde{Y}^{1},\widetilde{Y}^{2},\widetilde{Y}^{3})$ is an anti-realization of $\mathfrak{sl}(2,\mathbb{C})$.
It is immediate to check that, $\widetilde{Y}^{1},\widetilde{Y}^{2},\widetilde{Y}^{3}$ do not preserve the radius function $r^{2}=\delta_{jk}\,x^{j}\,x^{k}$ unless we are on the submanifold of pure quantum states ($r^{2}=1$):

\be
\widetilde{Y}^{j}(r^{2})=2(r^{2} - 1)x^{j}\,.
\ee
This means that the gradient-like vector fields are not tangent to the manifolds of isospectral quantum states, unless we consider pure quantum states.

\section{Lie groups, K\"{a}hler actions, and complexification}\label{sec: Lie groups, Kahler actions, and complexification}

Now, for a moment, we will switch our attention from the concrete case of isospectral orbits of $\SUh$ in $\mathfrak{T}_{1}(\Hh)$ to the more general framework of a K\"{a}hler manifold $(M,\,\omega,\,g,\,J)$ on which there is a Lie group $G$ acting in such a way as to preserve the K\"{a}hler structure of $M$.
In this context, we will prove that there is a canonical anti-realization of the complexification $\mathfrak{g}^{\mathbb{C}}$ of the Lie algebra $\mathfrak{g}$ of $G$ by means of vector fields on $M$.
In particular, if $M$ is compact, we obtain a left action of the complexification $G^{\mathbb{C}}$ of $G$ on $M$.
We will specialize to the case where $M\equiv\mathcal{O}$ is an orbit of $\SUh$ on $\mathfrak{T}_{1}(\Hh)$ (in particular, a manifold of isospectral quantum states) in the next subsection reobtaining the results presented in \cite{ciaglia_ibort_marmo-differential_geometry_of_quantum_states_observables_and_evolution} as a particular case of a more general instance.

\vsp

First of all, we recall the definition of the Nijenhuis tensor $N_{T}$ associated with a $(1,1)$ tensor field $T$ on a manifold $M$ (see definition $2.10$, and equation $2.4.26$ in \cite{marmo_ferrario_lovecchio_morandi_rubano-the_inverse_problem_in_the_calculus_of_variations_and_the_geometry_of_the_tangent_bundle}):

\be
N_{T}(X,Y)=\left(\mathcal{L}_{T(X)}(T)\right)(Y) - \left(T\circ \mathcal{L}_{X}(T)\right)(Y)\,,
\ee
where $X,Y$ are arbitrary vector fields on $M$.
A fundamental result in the theory of complex manifold is that  the $(1,1)$ tensor field defining the complex structure of a complex manifold must have vanishing Nijenhuis tensor \cite{nirenberg_newlander-complex_analytic_coordinates_in_almost_complex_manifolds}.
This means that, when $M$ is a K\"{a}hler manifold, the complex structure $J$ is such that $N_{J}=0$, which means:

\be
\left(\mathcal{L}_{J(X)}(J)\right)(Y) = \left(J \circ \mathcal{L}_{X}(J)\right)(Y)\,,
\ee
where $X,Y$ are arbitrary vector fields on $M$.
In particular, denoting by $X^{\mathbf{A}}$ the  fundamental vector field of the K\"{a}hlerian action of $G$ on $M$ associated with $\mathbf{A}\in\mathfrak{g}$, we have $\mathcal{L}_{X^{\mathbf{A}}}\,J=0$ because the action is K\"{a}hlerian, and thus:

\be\label{eqn: lie derivative of complex structure on Kahler manifold with respect to gradient vector fields of fundamental vector fields}
\left(\mathcal{L}_{Y^{\mathbf{A}}}(J)\right)(Z)=\left(\mathcal{L}_{J (X^{\mathbf{A}})}(J)\right)(Z) = 0
\ee
for every gradient vector field $Y^{\mathbf{A}}=J(X^{\mathbf{A}})$ on $\stsp_{\sigma}$.
Eventually, we can prove the following:

\begin{proposition}\label{prop: anti-realization of the complexification of the Lie algebra of a Lie group acting on a kahler manifold}
Let $G$ be a Lie group acting on the K\"{a}hler manifold $(M,\,\omega,\,g,\,J)$ by means of a left-action preserving the K\"{a}hler structure of $M$.
Let $\mathbf{A},\mathbf{B}$ be generic elements in the Lie algebra $\mathfrak{g}$ of $G$, then, the following commutation relations among the fundamental vector fields $X^{\mathbf{A}}$ of the action of $G$ on $M$ and their associated gradient vector fields $Y^{\mathbf{A}}=J(X^{\mathbf{A}})$ hold:

\be
[X^{\mathbf{A}}\,,X^{\mathbf{B}}]=-X^{[\mathbf{A}\,,\mathbf{B}]}\,,\;\;\;\;\;[X^{\mathbf{A}}\,,Y^{\mathbf{B}}]=- Y^{[\mathbf{A}\,,\mathbf{B}]}\,,\;\;\;\;\;[Y^{\mathbf{A}}\,,Y^{\mathbf{B}}]=X^{[\mathbf{A}\,,\mathbf{B}]}\,.
\ee

\begin{proof*}
The first commutator follows directly from the fact that the vector fields $X^{\mathbf{A}},X^{\mathbf{B}}$ are the fundamental vector fields of a left action of $G$.
Regarding the second commutator, we recall  equation \eqref{eqn: lie derivative of complex structure on Kahler manifold with respect to gradient vector fields of fundamental vector fields}, so that:

\be
\begin{split}
[X^{\mathbf{A}}\,,Y^{\mathbf{B}}]=&\mathcal{L}_{X^{\mathbf{A}}}\,\left(J(X^{\mathbf{B}})\right)= \\
=&\left(\mathcal{L}_{X^{\mathbf{A}}}\,J\right)(X^{\mathbf{B}}) + J\left(\mathcal{L}_{X^{\mathbf{A}}}\, X^{\mathbf{B}}\right)=\\
=& J\left([X^{\mathbf{A}}\,, X^{\mathbf{B}}]\right)=- Y^{ [\mathbf{A}\,,\mathbf{B}]}
\end{split}
\ee
as claimed.
Finally, using equation \eqref{eqn: lie derivative of complex structure on Kahler manifold with respect to gradient vector fields of fundamental vector fields} together with the fact that $J \circ J =-\mathrm{Id}$ because it is a complex structure, we obtain:

\be
\begin{split}
[Y^{\mathbf{A}}\,,Y^{\mathbf{B}}]=& \mathcal{L}_{J(X^{\mathbf{A}})}\,\left(J(X^{\mathbf{B}})\right)=\\
=& \left(\mathcal{L}_{J(X^{\mathbf{A}})}(J)\right)X^{\mathbf{B}} + J\left(\mathcal{L}_{J(X^{\mathbf{A}})}X^{\mathbf{B}}\right)=\\
=&J\left([Y^{\mathbf{A}}\,,X^{\mathbf{B}}]\right)=X^{[\mathbf{A}\,,\mathbf{B}]}
\end{split}
\ee 
as claimed.
\end{proof*}
\end{proposition}

From the commutators just computed we obtain an anti-realization of the Lie algebra $\mathfrak{g}^{\mathbb{C}}$ of the complexification $G^{\mathbb{C}}$ of $G$, and this anti-realization ``integrates'' to an action of $G^{\mathbb{C}}$ when every vector field $X^{\mathbf{A}} + Y^{\mathbf{B}}$ with $(\mathbf{A} + \imath\mathbf{B})\in\mathfrak{g}^{\mathbb{C}}$ is complete.
In particular, this is always the case if $M$ is a compact manifold.
Note that the gradient vector fields $Y^{\mathbf{A}}=J(X^{\mathbf{A}})$ need not preserve the symplectic structure because:

\be
\begin{split}
\left(\mathcal{L}_{Y^{\mathbf{A}}}\omega\right)(Y^{\mathbf{A}},\,X^{\mathbf{A}})\,&=\,\mathrm{d}\left(\mathrm{i}_{Y^{\mathbf{A}}}\omega\right)(Y^{\mathbf{A}},\,X^{\mathbf{A}})=\\
&=\,Y^{\mathbf{A}}\left(\omega(Y^{\mathbf{A}},\,X^{\mathbf{A}})\right) - X^{\mathbf{A}}\left(\omega(Y^{\mathbf{A}},\,Y^{\mathbf{A}})\right) - \omega(Y^{\mathbf{A}},\,[Y^{\mathbf{A}},\,X^{\mathbf{A}}])\,=\\
&=\,Y^{\mathbf{A}}\left(g(X^{\mathbf{A}},\,X^{\mathbf{A}})\right)
\end{split}
\ee
in general does not vanish.

\vsp

Now, let us come back to the case of finite-level quantum systems.
According to what has been previously reviewed, we have that the orbits of the action of $\SUh$ on the space $\mathfrak{T}_{1}(\Hh)$ of trace-one Hermitean operators are naturally compact K\"{a}hler manifolds on which $\SUh$ acts in such a way as to preserve the K\"{a}hler structure.
Consequently, proposition \ref{prop: anti-realization of the complexification of the Lie algebra of a Lie group acting on a kahler manifold} implies that the family of Hamiltonian and gradient vector fields associated with elements in the Lie algebra $\suh$ of the Lie group $\SUh$ provide an anti-realization of the Lie algebra $\slh$ which is the complexification of the Lie algebra $\suh$.
Furthermore, Hamiltonian and gradient vector fields are complete because the manifolds of isospectral quantum states are compact, and thus the anti-realization of $\slh$ ``integrates'' to a left action of $\SLh$ on every orbit.

In particular, when we consider the orbit corresponding to pure states (rank-one projectors in $\mathfrak{T}_{1}(\Hh)$), we are able to explicitely write out this left action of $\SLh$, and the result is precisely the action of $\SLh$ given in equation \eqref{eqn: action of SLh on quantum states}.
Note that the same is not true for quantum states with rank greater than $1$ because, in this case, the $\SLh$ action founded in this section preserves the spectrum, while the action given in equation \eqref{eqn: action of SLh on quantum states} does not.

According to remark \ref{rem: complex structure on pure quantum states}, the action of the complex structure $J$ on a vector field $X$ reads:

\be
\left(J(X)\right)(\rho)=[[\mathbf{a}\,,\rho]\,,\rho]=\{\mathbf{a}\,,\rho\} - 2 \rho\,\mathbf{a}\,\rho\,,
\ee
where $X(\rho)=\imath[\mathbf{a},\,\rho]$.
Consequently, the evaluation of the gradient vector field $Y^{\mathbf{A}}=J(X^{\mathbf{A}})$ at the pure quantum state $\rho$ may be written as:

\be
Y^{\mathbf{A}}(\rho)=J(X^{\mathbf{A}})(\rho)=[[\mathbf{a}\,,\rho]\,,\rho]\,,
\ee
where $\mathbf{A}=\imath\mathbf{a}$.
Now, let us consider the curve $\gamma_{\rho}^{\mathbf{A}}(t)$ starting at the pure quantum state $\rho$ and given by:

\be\label{eqn: isorank action of the boost part of GL(n,C)}
\gamma_{\rho}^{\mathbf{A}}(t):=\frac{\mathrm{e}^{t\mathbf{a}}\,\rho\,\mathrm{e}^{t\,\mathbf{a}}}{\Tr\left(\mathrm{e}^{t\mathbf{a}}\,\rho\,\mathrm{e}^{t\,\mathbf{a}}\right)}\,.
\ee
It is easy to see that $\gamma_{\rho}^{\mathbf{A}}(t)$ is always a pure quantum state, and that it satisfies the composition law:

\be\label{eqn: integral curve of gradient vector field on pure quantum states I}
\gamma_{\rho}^{\mathbf{A}}(t+s)=\gamma_{\rho_{s}}^{\mathbf{A}}(t)\,,\;\;\;\;\rho_{s}\equiv\gamma_{\rho}^{\mathbf{A}}(s)\,.
\ee
Then, denoting by $\mathbf{v}_{\rho}$  the tangent vector to $\gamma_{\rho}^{\mathbf{A}}(t)$ at $\rho$, a direct computation shows that:

\be\label{eqn: integral curve of gradient vector field on pure quantum states II}
\begin{split}
\left(\mathrm{d}e_{\mathbf{B}}(\rho)\right)(\mathbf{v}_{\rho})&=\frac{\mathrm{d}}{\mathrm{d}t}\,\left.e_{\mathbf{B}}\left(\gamma_{\rho}^{\mathbf{A}}(t)\right)\right|_{t=0}=\\
&=-\imath\,\frac{\mathrm{d}}{\mathrm{d}t}\,\left.\Tr\left(\mathbf{B}\,\frac{\mathrm{e}^{t\mathbf{a}}\,\rho\,\mathrm{e}^{t\,\mathbf{a}}}{\Tr\left(\mathrm{e}^{t\mathbf{a}}\,\rho\,\mathrm{e}^{t\,\mathbf{a}}\right)}\right)\right|_{t=0}=\\
&=-\imath\,\Tr\left(\mathbf{B}\,\{\mathbf{a},\,\rho\}\right) + 2\imath\,\Tr\left(\mathbf{B}\rho\right)\,\Tr\left(\mathbf{a}\rho\right)=\\
&=-\imath\,\Tr\left(\mathbf{B}\,\{\mathbf{a},\,\rho\}\right) + 2\imath\,\frac{\langle\psi_{\rho}|\mathbf{B}|\psi_{\rho}\rangle}{\langle\psi_{\rho}|\psi_{\rho}\rangle}\,\frac{\langle\psi_{\rho}|\mathbf{a}|\psi_{\rho}\rangle}{\langle\psi_{\rho}|\psi_{\rho}\rangle}=\\
&= -\imath\,\Tr\left(\mathbf{B}\,\{\mathbf{a},\,\rho\}\right) + 2\imath\,\Tr\left(\mathbf{B}\,\rho\,\mathbf{a}\rho\right)=\\
&=-\imath\,\Tr\left(\mathbf{B}\,[[\mathbf{a},\,\rho],\,\rho]\right)=\\
&=\left(\mathrm{d}e_{\mathbf{B}}(\rho)\right)\left(Y^{\mathbf{A}}(\rho)\right)\,,
\end{split}
\ee 
where we used the fact that every pure quantum state $\rho$ may be written as $\frac{|\psi_{\rho}\rangle\langle\psi_{\rho}|}{\langle\psi_{\rho}|\psi_{\rho}\rangle}$ for some nonzero $|\psi_{\rho}\rangle\in\Hh$.
Putting together equation \eqref{eqn: integral curve of gradient vector field on pure quantum states I} and equation \eqref{eqn: integral curve of gradient vector field on pure quantum states II} we conclude that $\gamma_{\rho}^{\mathbf{A}}(t)$ is the integral curve of $Y^{\mathbf{A}}$ starting at $\rho$.
Note that this result can not be generalized to quantum states that are not pure  because it is based on the equality:

\be
\Tr(\mathbf{A}\,\rho)=\frac{\langle\psi_{\rho}|\mathbf{A}|\psi_{\rho}\rangle}{\langle\psi_{\rho}|\psi_{\rho}\rangle}
\ee
which is true only for pure quantum states (rank-one projectors).

\vsp

It would be useful to be able to describe the isospectral action of $\SLh$ by means of vector fields on the affine hyperplane $\mathfrak{T}_{1}(\Hh)$ just as we did for the action of $\SLh$ preserving the rank.
A similar expression would be particularly useful in concrete cases because on $\mathfrak{T}_{1}(\Hh)$ we have global systems of Cartesian coordinates in which computations may be handled.

It turns out that it is ``almost'' possible to obtain the goal.
The basic idea is to introduce a $(1,1)$ tensor field $\mathscr{J}$ on $\mathfrak{T}_{1}(\Hh)$ which plays the role of the complex structure $J$ on the isospectral orbits, and define the gradient vector fields by applying $\mathscr{J}$ to the Hamiltonian vector fields $L^{\mathbf{A}}$ generating the action of $\SUh$ on $\mathfrak{T}_{1}(\Hh)$.
Clearly, $\mathscr{J}$ will not be a complex structure, and this has some consequences we must accept.

The $(1,\,1)$ tensor field $\mathscr{J}$ is defined by:

\be
\mathscr{J}\,:=\,\mathcal{N}\,\circ\,\Lambda\,,
\ee
where $\mathcal{N}$ is the Riemannian metric on $\mathfrak{T}_{1}(\Hh)$ which is associated with the symmetric part of the Hilbert inner product on $\Bh$ (see equation \eqref{eqn: Riemannian metric tensor on affine hyperplane}).
In coordinates we have:

\be
\mathscr{J}\,=\,\sum_{j}\,\mathrm{d}x^{j}\,\otimes\,L^{j}\,=\,\sum_{j}\,c^{kj}_{l}\,x^{l}\,\mathrm{d}x^{j}\,\otimes\,\frac{\partial}{\partial x^{k}}\,.
\ee
Since $\mathcal{N}$ and $\Lambda$ are invariant with respect to the action of $\SUh$ on $\mathfrak{T}_{1}(\Hh)$ generated by the $L^{\mathbf{A}}$'s, that is:

\be
\mathcal{L}_{L^{\mathbf{A}}}\,\Lambda\,=\,\mathcal{L}_{L^{\mathbf{A}}}\,\mathcal{N}\,=\,0\,,
\ee
we have that $\mathscr{J}$ is invariant too, that is:

\be
\mathcal{L}_{L^{\mathbf{A}}}\,\mathscr{J}\,=\,0\,.
\ee
Furthermore, by direct inspection we find that the vector field:

\be
\widehat{Y}_{f}\,=\,\mathscr{J}(X_{f})\,,
\ee
where $X_{f}$ is the Hamiltonian vector field associated with the smooth function $f$ on $\mathfrak{T}_{1}(\Hh)$ by means of $\Lambda$, is always tangent to the isospectral orbits of $\SUh$ on $\mathfrak{T}_{1}(\Hh)$.
Indeed:

\be
Y_{f}\,=\,\sum_{j}\,\mathrm{d}x^{j}(X_{f})\,L^{j}\,,
\ee
which means that every $\widehat{Y}_{f}$ is written in terms of the $L^{j}$, and the $L^{j}$'s are always tangent to the isospectral orbits.
Now, we may introduce a contravarian tensor $\mathscr{G}$ on $\mathfrak{T}_{1}(\Hh)$ by setting:

\be
\mathscr{G}\,:=\,\mathscr{J}\,\circ\,\Lambda\,=\,\delta_{jk}\,L^{j}\,\otimes\,L^{k}\,.
\ee
Clearly, $\mathscr{G}$ is invariant with respect to the action of $\SUh$ on $\mathfrak{T}_{1}(\Hh)$, and it is clear that the ``restriction'' $\mathcal{G}\equiv\left.\mathscr{G}\right|_{\mathcal{O}}$ of $\mathscr{G}$ to any isospectral orbit $\mathcal{O}$ is invertible.
Then, we may consider the inverse $\mathcal{G}^{-1}$ which is a Riemannian metric on $\mathcal{O}$ which is invariant with respect to the canonical action of $\SUh$ on $\mathcal{O}$.
The invariance of $\mathcal{G}^{-1}$ with respect to the action of $\SUh$ on the isospectral orbit $\mathcal{O}$ together with the fact that $\SUh$ is a simple Lie group force $\mathcal{G}^{-1}$ to be a constant multiple of the metric tensor $g$ associated with the canonical K\"{a}hler structure on $\mathcal{O}$.
Clearly, the constant depends on the orbit $\mathcal{O}$ we are considering.

Now, we may write $\widehat{Y}_{f}$ as the ``gradient'' vector field associated with $f$ by means of $\mathscr{G}$:

\be
\widehat{Y}_{f}\,=\,\mathscr{J}(X_{f})\,=\,\mathscr{J}\,\circ\,\Lambda(\mathrm{d}f)\,=\,\mathscr{G}(\mathrm{d}f)\,,
\ee
and thus we see that the restriction of $\widehat{Y}_{f_{\mathbf{A}}}\equiv \widehat{Y}^{\mathbf{A}}$ to $\mathcal{O}$ is a constant multiple of the gradient vector field $Y^{\mathbf{A}}$ introduced before.
As said before, the constant relating the restriction of $\widehat{Y}^{\mathbf{A}}$ to $Y^{\mathbf{A}}$ depends on the isospectral orbit $\mathcal{O}$ we are considering.
Unfortunately, we can not find a smooth function $F$ such that the restriction of $F\,\widehat{Y}^{\mathbf{A}}$ to $\mathcal{O}$ is precisely $Y^{\mathbf{A}}$ for every $\mathcal{O}$  unless we consider the q-bit case (in which case we must exclude the maximally mixed state from the game).

In the end, it is possible to describe the infinitesimal action of $\SLh$ on isospectral orbits by means of vector fields on the affine hyperplane $\mathfrak{T}_{1}(\Hh)$ as long as we accept that the gradient part on $\mathfrak{T}_{1}(\Hh)$ is only ``conformally related'' with the gradient part on the isospectral orbits, and the conformal factor depends on the isospectral orbit in a non-smooth manner.

\begin{remark}
It is well-known that the Hamiltonian vector fields $L^{\mathbf{A}}$ give rise to linear transformations that preserve the spectrum, and thus the von Neumann entropy:

\be
\mathit{S}_{vN}(\rho)\,=\,\mathrm{Tr}\left(\rho\,\ln(\rho)\right)\,.
\ee
According to the results presented in this section, we see that the gradient vector fields  $\widehat{Y}^{\mathbf{A}}$   give rise to nonlinear transformations that preserve the spectrum of quantum states and thus preserve the von Neumann entropy, while the gradient-like vector fields $\widetilde{Y}^{\mathbf{A}}$ introduced in section \ref{sec: Quantum states and the special unitary group} give rise to nonlinear transformations that do not preserve the spectrum of quantum states and thus do not preserve von Neumann's entropy.
\end{remark}

\section{The q-bit II}\label{sec: q-bit II}

Let us now look again at what happens to the structures introduced in section \ref{sec: Lie groups, Kahler actions, and complexification} in the concrete case of a q-bit, that is, when $\mathcal{H}\cong\mathbb{C}^{2}$.
Referring to the notation of section \ref{sec: q-bit I} and writing $\widehat{Y}^{j}\equiv\widehat{Y}_{f_{j}}$ with $f_{j}(\xi)=x^{j}(\xi)$, we have:

\be
\begin{split}
\mathscr{G}\,&=\,\delta_{jk}\,L^{j}\,\otimes\,L^{k}\,=\\
&\,=4\delta_{jk}\epsilon^{js}_{l}\,\epsilon^{kt}{m}\,x^{l}\,x^{m}\,\frac{\partial}{\partial x^{s}}\,\otimes\,\frac{\partial }{\partial x^{t}}\,=\\ 
&=\,4\left(r^{2}\,\delta_{jk}\,\frac{\partial}{\partial x^{j}}\,\otimes\,\frac{\partial}{\partial x^{k}} - \Delta\,\otimes\,\Delta\right)\,,
\end{split}
\ee
where $r^{2}=\delta_{ls}x^{l}x^{s}$ and $\Delta=x^{k}\,\frac{\partial}{\partial x^{k}}$ is the dilation vector field, and thus:

\be
\widehat{Y}^{j}\,=\,\mathscr{G}(\mathrm{d}x^{j})\,=\,4\left(r^{2}\,\frac{\partial}{\partial x^{j}} - x^{j}\Delta\right)\,.
\ee
Concerning the $(1,1)$ tensor field $\mathscr{J}$ of equation \eqref{eqn: quasi-complex structure}, we note that, in the case of the q-bit, all the isospectral orbits are diffeomorphic to 2-dimensional spheres, except the degenerate orbit passing through the maximally mixed state, that is, the point with $(x^{1},\,x^{2},\,x^{3})=(0,\,0,\,0)$.
Consequently, removing the maximally mixed state $\rho_{m}$ from $\traceh$, the constant relating the restriction to the isospectral orbit $\mathcal{O}$ of $\widehat{Y}_{j}$  with the gradient vector field $Y_{j}$ on the isospectral orbit $\mathcal{O}$ is essentially given by the restriction of the radius function $r^{2}=\delta_{jk}x^{j}x^{k}$ to $\mathcal{O}$.
Now, we introduce the $(1,1)$ tensor field:

\be
\mathfrak{J}\,:=\,\frac{1}{2r}\,\mathscr{J}\,=\,\sum_{j}\,\frac{1}{2r}\,\mathrm{d}x^{j}\,\otimes\,L^{j}\,,
\ee
and note that it is invariant with respect to the action of the special unitary group generated by the Hamiltonian vector fields $(L^{1},\,L^{2},\,L^{3})$, and that:

\be
\mathfrak{J}^{3}\,=\,-\mathfrak{J}\,.
\ee
By means of this last property, we may look at $\mathfrak{J}$ as a generalization of a complex structure to an odd-dimensional manifold.
Indeed, every complex structure $J$ is such that $J^{2}=-\mathrm{Id}$ and thus it is such that $J^{3}=-J$, however, this latter conditions may be implemented in the case of odd-dimensional manifolds while $J^{2}=-\mathrm{Id}$ require the manifold to be even-dimensional.
Now, we may build the pseudo-gradient vector fields on $(\traceh - \{\rho_{m}\})$:

\be
Y^{j}\,:=\,\mathfrak{J}(L^{j})\,=\,\frac{2}{r}\left(r^{2}\,\frac{\partial}{\partial x^{j}} - x^{j}\Delta\right)\,.
\ee
It is a matter of direct computation to show that the $Y^{j}r^{2}=0$ for all $j=1,2,3$, that is, $Y^{j}$ is tangent to every isospectral orbit, and that $(L^{1},\,L^{2},\,L^{3},\,Y^{1},\,Y^{2},\,Y^{3})$ close on a realization of $\SLh$.

Having $\mathfrak{J}$, we may compose it with $\Lambda$ in order to obtain the symmetric bivector field:

\be
G=\mathfrak{J}\circ\Lambda=\frac{1}{2r}\,\mathscr{G}\,=\,\frac{2}{r}\left(r^{2}\delta^{jk}\,\frac{\partial}{\partial x^{j}}\otimes\,\frac{\partial}{\partial x^{k}}- \Delta\otimes\Delta\right)\,.
\ee
Clearly, it is $Y^{j}=G(\mathrm{d}x^{j})$ by construction, and $G$ resembles the (contravariant form of the negative of the) Killing form for $\SUh$ divided by the Casimir function $2r$.
What is interesting is that we may start with the symmetric bivector $\mathcal{R}$ and compose it with $\mathfrak{J}$ to obtain:

\be
\mathfrak{J}\circ\mathcal{R}=\frac{1}{4r}\,\Lambda\,.
\ee
This means that the Hamiltonian vector fields $L^{\mathbf{A}}$ may be written as $4r\,\mathfrak{J}(\widetilde{Y}^{\mathbf{A}})$, where $\widetilde{Y}^{\mathbf{A}}$ are the gradient-like vector fields.

Consequently, we see that the different ways in which $\SLh$ acts on $\stsp$ come from the interplay between the Poisson bivector field $\Lambda$ and the $(1,1)$-tensor field $\mathfrak{J}$.
We obtain the action of $\SLh$ on the manifolds of isospectral quantum states using the Hamiltonian vector fields $L^{\mathbf{A}}$, with $\mathbf{A}\in\suh$, and the pseudo-gradient vector fields $Y^{\mathbf{A}}:=\mathfrak{J}(L^{\mathbf{A}})$, while,  the action of $\SLh$ on the manifolds of quantum states with fixed rank is generated by the Hamiltonian vector fields $L^{\mathbf{A}}$, with $\mathbf{A}\in\suh$, and the gradient-like vector fields $\widetilde{Y}^{\mathbf{A}}$ defined requiring that $L^{\mathbf{A}}=4r\,\mathfrak{J}(\widetilde{Y}^{\mathbf{A}})$.
The fact that $\Lambda$ and $\mathfrak{J}$ are not invertible is responsible for the fact that the pseudo-gradient vector fields are different from the gradient-like vector fields, and thus, for the fact that $\SLh$ acts on the space $\stsp$ of quantum states in two different ways.

\begin{remark}
The possibility of describing the isospectral action of $\SLh$ in terms of the the Hamiltonian vector fields $L^{\mathbf{A}}$ and their images through a $(1,1)$ tensor field $\mathfrak{J}$ such that $\mathfrak{J}^{3}=-\mathfrak{J}$ on (some open submanifold of) $\mathfrak{T}_{1}(\mathcal{H})$ in higher dimensional cases requires a deeper analysis, and will be dealt with in future works.
\end{remark}

\section{Composite systems}\label{sec: composite systems}

Composite quantum systems and quantum entanglement manipulation are of fundamental importance in the context of quantum information theory and the geometry of quantum entanglement is a fascinating and very complex subject \cite{aniello_clemente-gallardo_marmo_volkert-classical_tensors_and_quantum_entanglement_I, bengtsson_zyczkowski-geometry_of_quantum_states:_an_introduction_to_quantum_entanglement, chirco_mele_oriti_vitale-fisher_metri_geometric_entanglement_and_spin_networks, grabowski_kus_marmo-geometry_of_quantum_systems_density_states_and_entanglement, grabowski_kus_marmo-symmetries_group_actions_and_entanglement, gurvits_barnum-separable_balls_around_the_maximally_mixed_multipartite_quantum_state, horodecki-quantum_entanglement, huckleberry_kus_sawicki-bipartite_entanglement_spherical_ations_and_geometry_of_local_unitary_orbits,  leinaas_myrheim_ovrum-geometrical_aspects_of_entanglement, samuel_shivam_sinha-lorentzian_geometry_of_qubit_entanglement, sanpera_bruss_lewenstein-schmidt-number_witnesses_and_bound_entanglement, huckleberry_kus_sawicki-symplectic_geometry_of_entanglement, kus_oszmaniec_sawicki-convexity_of_momentum_map_morse_index_and_quantum_entanglement, terhal_horodecki-schmidt_number_for_density_matrices, verstraete_audenaert_demoor-maximally_entangled_mixed_states_of_two_qubits, zyczkowski_horodecki_sanpera_lewenstein-volume_of_the_set_of_separable_states}.
Here we want to present some possible connections between the geometry of quantum entanglement and the geometrical tools introduced in the previous sections. 

At this purpose, let us start considering a bipartite quantum systems made of two q-bits.
In this case, it has been recently proposed to introduce a local action of the Lorentz group in order to detect quantum entanglement of non normalized quantum states, that is, positive semi-definite linear operators on the Hilbert space of two q-bits \cite{leinaas_myrheim_ovrum-geometrical_aspects_of_entanglement, samuel_shivam_sinha-lorentzian_geometry_of_qubit_entanglement}.
Specifically, consider a positive semi-definite operator $\mathbf{P}$ of a two-level quantum system as an element $(x^{0},\,x^{1},\,x^{2},\,x^{3})$ in the Minkowski spacetime $\mathcal{M}$ according to the identification:

\be
\mathbf{P}=P^{\mu}\,\sigma_{\mu}\;\;\longleftrightarrow\;\;(x^{0}\equiv P^{0},\,x^{1}\equiv P^{1},\,x^{2}\equiv P^{2},\,x^{3}\equiv P^{3})\,.
\ee
Then, the positivity condition on $\mathbf{P}$ may be rephrased in a relativistic language as $g_{\mu\nu}\,x^{\mu}\,x^{\nu}\geq 0$.
Projectors (pure non-normalized states) are identified with light-like vectors, while non-normalized mixed states are identified with time-like vectors.
Clearly, the causal-like vector $(x^{0},\,x^{1},\,x^{2},\,x^{3})$ is future-pointing because $x^{0}=\Tr(\mathbf{P})$ is positive.
In this setting, it is immediate to realize that the canonical action $\mathbf{x}\rightarrow L(\mathbf{x})$ of the (proper orthocronus) Lorentz group on $\mathcal{M}$ preserves the positive character of $\mathbf{P}$.
When two q-bits are considered, we obtain two local actions of the (proper orthocronus) Lorentz group on each of the two subsystems, and it is clear that these actions preserve separability (entanglement). 
This observation allows the authors to use the mathematical tools of relativity theory (e.g., the Dominant Energy Condition and the Strong Energy Condition) in order to provide a necessary and sufficient separability test for a positive semi-definite operator $\mathbf{P}$, as well as an algorithm for constructing the separable form of a given separable positive semi-definite operator.

It is clear that the results of \cite{samuel_shivam_sinha-lorentzian_geometry_of_qubit_entanglement} depend on the fact that the Lorentz group ``naturally acts on (tensor products of) q-bits'', while this is no longer true for a generic $n$-level quantum sytem.
Indeed, at the end of the papers, the authors explicitely states that in order to generalize their results to the higher dimensional case it is necessary to find the correct (maximal) group of separability-preserving transformation.

We feel that the geometric setting discussed in the present work could point towards the definition of such group.
Indeed, the canonical realization of the (proper orthocronus) Lorentz group on $\mathcal{M}$ may be represented in the space of positive semi-definite operators on the Hilbert space of a q-bit by means of the action of the special linear group $\SLh = SL(2,\,\mathbb{C})$:

\be
\mathbf{P}\;\;\rightarrow\;\;\gr\,\mathbf{P}\,\gr^{\dagger}\,,\;\;\;\;\gr\in\SLh = SL(2,\,\mathbb{C})\,.
\ee
Then, the transition from the (proper orthocronus) Lorentz group to the special linear group $\SLh = SL(2,\,\mathbb{C})$ allows us to immediately realize that the generalization to the higher dimensional case requires the special linear group $\SLh$.
A similar solution is proposed in \cite{leinaas_myrheim_ovrum-geometrical_aspects_of_entanglement}, where, however, the normalization condition $\Tr(\rho)=1$ for a quantum state $\rho$ is relaxed in order to define an action of $\SLh$.
According to what has been said in the present work, we are now able to take into account the normalization condition $\Tr(\rho)=1$ of quantum states defining an action of $\SLh$ directly on the space $\stsp$ of quantum states preserving this constraint.

\vsp

Let us now review some general aspects of the interplay between entanglement features and the geometry of quantum states.
Consider a composite quantum system living in $\mathcal{H} = \mathcal{H}_A \otimes\mathcal{H}_B$. 
Fixing an orthonormal basis $\{e_1,\ldots,e_{n_A}\}$ in  $\mathcal{H}_A$, and $\{f_1,\ldots,f_{n_B}\}$ in  $\mathcal{H}_B$, any vector $\psi \in \mathcal{H}$ may be represented as follows

\begin{equation}\label{}
  \psi = \sum_{i=1}^{n_A} \sum_{j=1}^{n_B} \, \psi_{ij} e_i \otimes f_j .
\end{equation}
Equivalently, one has

\begin{equation}\label{}
  \psi =  \sum_{i=1}^{n_A}  e_i \otimes \hat{\psi} e_i ,
\end{equation}
where the linear operator $\hat{\psi} : \mathcal{H}_A \to \mathcal{H}_B$ is defined via

\begin{equation}\label{}
  \hat{\psi}e_i = \sum_{j=1}^{n_B} \, \psi_{ij} f_j .
\end{equation}
Note that $\psi$ is normalized iff ${\rm Tr}(\hat{\psi}^\dagger \hat{\psi})=1$. 
One calls the rank of $\hat{\psi}$ a Schmidt rank of $\psi$: ${\rm SR}(\psi) := {\rm rank}(\hat{\psi})$. Clearly, $\psi$ is separable iff ${\rm SR}(\psi)=1$. This definition may be generalized to density operators representing quantum states as follows: any density operator can be represented as a convex combination of rank-1 projectors (not necessarily mutually orthogonal)

\begin{equation}\label{}
  \rho = \sum_i p_i P_i ,
\end{equation}
where $p_i$ is a probability distribution, and $P_i$ is a rank-1 projector corresponding to a vector $\psi_i \in \mathcal{H}$. Now, one defines  a Schmidt number of $\rho$ as follows \cite{sanpera_bruss_lewenstein-schmidt-number_witnesses_and_bound_entanglement, terhal_horodecki-schmidt_number_for_density_matrices}:

\begin{equation}\label{SN}
  {\rm SN}(\rho) := \min_{p_k,\psi_k} \{ \max_k\, {\rm SR}(\psi_k) \} .
\end{equation}
Again, $\rho$ is separable iff ${\rm SN}(\rho)=1$. 
In this case one can always find the following well known representation
\begin{equation}\label{}
  \rho = \sum_{j} \pi_{j} \rho^{j}_{A} \otimes \rho^{j}_{B} ,
\end{equation}
where $\rho^{j}_{A}$ and $\rho^{j}_{B}$ are density operators on $\mathcal{H}_{A}$ and $\mathcal{H}_{B}$, respectively, and $\pi_{j}$ is a probability distribution. It should be stressed that the Schmidt number defined in (\ref{SN}) is hardly computable and the problem to decide whether given $\rho$ represents a separable state is NP-hard. 
Writing:

\begin{equation}\label{}
  \mathcal{P}_k(\mathcal{H}) = \{ \rho \, |\, {\rm SN}(\rho) \leq k \}
\end{equation}
one has the following chain of inclusions

\begin{equation}\label{PPP}
  \mathcal{P}_{\rm sep}(\mathcal{H}) :=  \mathcal{P}_1(\mathcal{H})\subset  \mathcal{P}_2(\mathcal{H}) \subset \ldots \subset  \mathcal{P}_n(\mathcal{H}) = S
\end{equation}
where $n=\min\{n_A,n_B\}$. 

The notion of a Schmidt number is not compatible with the {\em global} action of $\SUh$ and $\SLh$, that is, in general
 $ {\rm SN}(U\rho U^\dagger) \neq {\rm SN}(\rho)$, and ${\rm SN}(g\rho g^\dagger/{\rm Tr}(g\rho g^\dagger) ) \neq {\rm SN}(\rho)$
for $U \in \SUh$ and $g \in \SLh$. However, 
for the   {\em local} actions $\mathcal{SU}(\mathcal{H}_A) \otimes \mathcal{SU}(\mathcal{H}_B)$ and $\mathcal{SL}(\mathcal{H}_A) \otimes \mathcal{SL}(\mathcal{H}_B)$ one has:

\begin{equation}\label{}
  {\rm SN}(U_A \otimes U_B \,\rho\, U_A^\dagger \otimes U_B^\dagger) = {\rm SN}(\rho)  ,
\end{equation}
and 

\begin{equation}\label{}
  {\rm SN}\left( \frac{\gr_{A} \otimes \gr_{B} \,\rho\, g_{A}^\dagger\otimes \gr_{B}^\dagger}{{\rm Tr}(\gr_{A} \otimes \gr_{B}\rho \gr_{A}^\dagger\otimes \gr_{B}^\dagger )} \right) = {\rm SN}(\rho) .
\end{equation}

Now, let $\mathcal{O}$ be an orbit of $\SUh$ passing through a quantum state $\rho$. 
The local action of $\mathcal{SU}(\mathcal{H}_A) \otimes \mathcal{SU}(\mathcal{H}_B)$ on $\mathcal{O}$ that preserves the K\"ahler structure and hence Hamiltonian and gradient vector fields  associated with elements from $\mathfrak{su}(\mathcal{H}_A) \otimes \mathfrak{su}(\mathcal{H}_B)$ provide anti-realization of the Lie algebra $\mathfrak{sl}(\mathcal{H}_A) \otimes \mathfrak{sl}(\mathcal{H}_B)$. 
Again, since $\mathcal{O}$ is compact, the Hamiltonian and gradient vector fields are complete and thus the anti-realization of $\mathfrak{sl}(\mathcal{H}_A) \otimes \mathfrak{sl}(\mathcal{H}_B)$ ``integrates" to a left action of $\mathcal{SL}(\mathcal{H}_A) \times \mathcal{SL}(\mathcal{H}_B)$ on every orbit of isospectral states (see section \ref{sec: Lie groups, Kahler actions, and complexification}). 
In particular for the orbit of pure states the corresponding Hamiltonian vector field $X^{\mathbf{H}_A\otimes \mathbf{H}_B}(\rho) = i[\mathbf{h}_A \otimes \mathbf{h}_B,\rho]$, and

\begin{equation}\label{}
  Y^{\mathbf{H}_A\otimes \mathbf{H}_B}(\rho) = (J(X^{\mathbf{H}_A\otimes \mathbf{H}_B}))(\rho) =[[\mathbf{h}_A \otimes \mathbf{h}_B,\rho],\rho] ,
\end{equation}
with $\mathbf{H}_\alpha = i\mathbf{h}_\alpha$ $(\alpha=A,B)$. 
The integral curves of $X^{\mathbf{H}_A\otimes \mathbf{H}_B}$ and $Y^{\mathbf{H}_A\otimes \mathbf{H}_B}$ consist of isospectral states with the same Schmidt number. 

It would be interesting to find spectral conditions which guarantee that all isospectral states belong to certain $\mathcal{P}_k(\mathcal{H})$ with $k< n$.
For instance, it could be useful to consider the local version of the isospectral action of $\SLh$ on the manifolds of isospectral quantum states defined in section \ref{sec: Lie groups, Kahler actions, and complexification} in order to analyze entanglement-invariant geometrical quantities connected with the symmetries and the constants of the motion of the Hamiltonian and gradient vector fields generating the local action.
We will pursue this ``mechanical point of view'' on quantum entanglement in a future work.

\vsp

Concerning the local action of $\mathcal{SL}(\Hh_{A})\otimes\mathcal{SL}(\Hh_{B})$ that preserves the rank (see equation \eqref{eqn: action of SLh on quantum states}), we already saw that it preserves the separability (or entanglement) of quantum states.
Then, if $\rho_{AB}=\rho_{A}\otimes\rho_{B}$ denotes a product state in $\stsp_{k}(\Hh)$ and $\gr_{A}\otimes\gr_{B}$ is an element in $\mathcal{SL}(\Hh_{A})\otimes\mathcal{SL}(\Hh_{B})$, we have that the action of $\gr_{A}\otimes\gr_{B}$ on $\rho_{AB}$ given by equation \eqref{eqn: action of SLh on quantum states} reads:

\be\label{eqn: local action of local special linear group on bipartite systems}
\frac{\gr_{A}\otimes\gr_{B}\,\rho_{AB}\,\gr_{A}^{\dagger}\otimes\gr_{B}^{\dagger}}{\mathrm{Tr}\left(\gr_{A}\otimes\gr_{B}\,\rho_{AB}\,\gr_{A}^{\dagger}\otimes\gr_{B}^{\dagger}\right)}\,=\,\left(\frac{\gr_{A}\rho_{A}\gr_{A}^{\dagger}}{\mathrm{Tr}_{A}(\gr_{A}\rho_{A}\gr_{A}^{\dagger})}\right)\otimes\left(\frac{\gr_{B}\rho_{B}\gr_{B}^{\dagger}}{\mathrm{Tr}_{B}(\gr_{B}\rho_{B}\gr_{B}^{\dagger})}\right)
\ee
which is again a product state in $\stsp_{k}(\Hh)$.
In particular, we may introduce the $(A,B)$-rank for a product state $\rho_{AB}=\rho_{A}\otimes\rho_{B}$ as the  couple of integer numbers $(\mathrm{rk}(\rho_{A}),\,\mathrm{rk}(\rho_{B}))$ with $\mathrm{rk}(\rho_{A})$ and $\mathrm{rk}(\rho_{B})$ being the rank of $\rho_{A}$ and $\rho_{B}$ as linear operators on $\Hh_{A}$ and $\Hh_{B}$ respectively.
Note that two product states with the same $(A,B)$-rank  must necessarily have the same rank as linear operators on $\Hh$, while product states with the same rank may have different $(A,B)$-rank.
A moment of reflection allows to conclude that the local action of $\mathcal{SL}(\Hh_{A})\otimes\mathcal{SL}(\Hh_{B})$  given by equation \eqref{eqn: local action of local special linear group on bipartite systems} is transitive on every subset of product states with fixed $(A,B)$-rank, and this means that product states  with fixed $(A,B)$-rank can be given the structure of homogeneous spaces for $\mathcal{SL}(\Hh_{A})\times\mathcal{SL}(\Hh_{B})$.
It is then easy to generalize these results to the case of a multipartite quantum system of distinguishable components.

We can take one step further, and consider all those separable states $\rho_{K}$ for which we can find a minimal decomposition in terms of $K$ product states:

\be
\rho_{K}\,=\,\sum_{j=1}^{K}\,\pi_{j}\,\rho^{j}_{A}\,\otimes\rho_{B}^{j}\,.
\ee
Minimality here refers to the fact that it is not possible to write $\rho_{K}$ as a convex combination of $R$ product states with $R<K$.
We call these states $K$-decomposable.
In this setting, product states are a particular case, namely, they are $1$-decomposable quantum states.
It is clear that the local action of $\mathcal{SL}(\Hh_{A})\otimes\mathcal{SL}(\Hh_{B})$ can not increase $K$-decomposability, indeed:

\be
\frac{\gr_{A}\otimes\gr_{B}\,\rho_{K}\,\gr_{A}^{\dagger}\otimes\gr_{B}^{\dagger}}{\mathrm{Tr}\left(\gr_{A}\otimes\gr_{B}\,\rho_{AB}\,\gr_{A}^{\dagger}\otimes\gr_{B}^{\dagger}\right)}\,=\,\sum_{j=1}^{K}\,\pi_{j}\,\left(\frac{\gr_{A}\rho_{A}^{j}\gr_{A}^{\dagger}}{\mathrm{Tr}_{A}(\gr_{A}\rho_{A}^{j}\gr_{A}^{\dagger})}\right)\otimes\left(\frac{\gr_{B}\rho_{B}^{j}\gr_{B}^{\dagger}}{\mathrm{Tr}_{B}(\gr_{B}\rho_{B}^{j}\gr_{B}^{\dagger})}\right)\,.
\ee
On the other hand, suppose that the local action of $\mathcal{SL}(\Hh_{A})\otimes\mathcal{SL}(\Hh_{B})$ decreases $K$-decomposability, that is, it changes $K$ to be $R<K$, and write:

\be
\rho_{R}^{\gr_{A}\otimes\gr_{B}}\,:=\,\frac{\gr_{A}\otimes\gr_{B}\,\rho_{K}\,\gr_{A}^{\dagger}\otimes\gr_{B}^{\dagger}}{\mathrm{Tr}\left(\gr_{A}\otimes\gr_{B}\,\rho_{AB}\,\gr_{A}^{\dagger}\otimes\gr_{B}^{\dagger}\right)}\,.
\ee
Then, if we apply the local action of $(\gr_{A}\otimes\gr_{B})^{-1}$ on $\rho_{R}^{\gr_{A}\otimes\gr_{B}}$, it is clear that we get back $\rho_{K}$, and thus $R$ changes in $K>R$, but this is not possible, and we conclude that the local action of $\mathcal{SL}(\Hh_{A})\times\mathcal{SL}(\Hh_{B})$ preserves $K$-decomposability.
Unfortunately, the local action of $\mathcal{SL}(\Hh_{A})\times\mathcal{SL}(\Hh_{B})$ is not transitive on the sets of $K$-decomposable quantum states unless $K=1$.

\vsp

It is important to note that there exist orbits of the action of $\SUh$ consisting of separable states only. 
It was proved in \cite{gurvits_barnum-separable_balls_around_the_maximally_mixed_multipartite_quantum_state, zyczkowski_horodecki_sanpera_lewenstein-volume_of_the_set_of_separable_states} that if $\rho$ satisfies
\begin{equation}\label{super}
  {\rm Tr} \rho^2 \leq \frac{1}{N-1} ,
\end{equation}
where $N=n_{A}n_{B}$, then all isospectral states are separable. 
Recall, that the purity is lower bounded by $1/N$ which shows that  $\rho$ satisfying (\ref{super}) has very low purity. 

\begin{remark}

For two qubit case $n_A=n_B=2$ it was proved in \cite{verstraete_audenaert_demoor-maximally_entangled_mixed_states_of_two_qubits} that if $\lambda_1\geq \lambda_2 \geq \lambda_3 \geq \lambda_4$ are eigenvalues of $\rho$ satisfying

\begin{equation}\label{}
  \lambda_1 - \lambda_3 \leq 2 \sqrt{\lambda_2 \lambda_4} ,
\end{equation}
then all isospectral states are separable.
Other examples of such conditions were found in \cite{gabach-clement_raggio-a_simple_spectral_condition_implying_separability_for_states_of_bipartite_quantum_systems}, specifically, for two qubits we have that if 

\begin{equation}\label{}
  3\lambda_1 + \sqrt{2}\lambda_2 + (3-\sqrt{2}) \lambda_3 \leq 0 ,
\end{equation}
then all isospectral states are separable. 
Moreover, if $\lambda_1 \geq \ldots \geq \lambda_N$ together with
\begin{equation}\label{}
  3\lambda_N + (N-1) \lambda_{N-1} \leq 0 ,
\end{equation}
the again all isospectral states are separable. 
\end{remark}

From the point of view of the affine hyperplane $\mathfrak{T}_{1}(\Hh)$, we have that $\rho$ is separable whenever it lies inside the sphere $S_{R}$ centered at the maximally mixed state $\rho_{m}\,=\,\frac{\mathbb{I}}{N}$ with radius $R=\sqrt{\frac{1}{N(N-1)}}$.
In particular, according to remark \ref{rem: lower bound on purity depends on rank}, we conclude that all the separable states inside the sphere $S_{R}$ must have maximal rank.

\section{Conclusions}\label{sec: conclusions}

The subject of this paper is the study of the geometrical structures adapted to the space $\stsp$ of quantum states of a finite-dimensional quantum system.
The space $\stsp$ is a convex body in the affine hyperplane $\mathfrak{T}_{1}(\Hh)$ of Hermitean operators on the Hilbert space $\Hh$ of the system having trace equal to $1$.
In particular, there is no global differential structure on $\stsp$, and we have to exploit the differential structure of the ambient space $\mathfrak{T}_{1}(\Hh)$ in order to describe geometrical objects related to $\stsp$.

In section \ref{sec: Quantum states and the special unitary group} we analyze two different partitions of $\stsp$ by means of group actions.
On the one hand, it is well known that the special unitary group $\SUh$ acts on $\stsp$ partitioning it into the disjoint union of manifolds of isospectral quantum states, and these manifolds are compact smooth embedded submanifolds of $\mathfrak{T}_{1}(\Hh)$. 
On the other hand, there is the action of another Lie group, namely, the special linear group $\SLh$ (the complexification of $\SUh$), which gives a coarser partition of $\stsp$ into the disjoint union of manifolds of quantum states with the same rank.
The group actions giving rise to these two stratifications may be described, infinitesimally, by means of Hamiltonian and gradient-like vector fields.
Regarding the action of the special unitary group $\SUh$, there is a canonical Poisson bivector $\Lambda$ on $\mathfrak{T}_{1}(\Hh)$ coming from the Lie algebra structure of $\suh$.
Specifically, denoting by $f_{\mathbf{A}},f_{\mathbf{B}}$ the linear functions on $\mathfrak{T}_{1}(\Hh)$ associated with $\mathbf{A},\mathbf{B}\in\uh$, we have:

\be
\Lambda(\mathrm{d}f_{\mathbf{A}},\,\mathrm{d}f_{\mathbf{B}}) := f_{\mathbf{A}\mathbf{B} - \mathbf{B}\mathbf{A}}\,,
\ee
and the Hamiltonian vector fields associated with the linear functions on $\mathfrak{T}_{1}(\Hh)$ are the fundamental vector fields of the action of $\SUh$.
These vector fields are denoted by $L^{\mathbf{A}}$, with $\mathbf{A}\in\suh$.
From the point of view of the theory of Poisson manifolds, the submanifolds of isospectral quantum states are the symplectic leaves of the symplectic foliation associated with the Poisson bivector $\Lambda$ \cite{weinstein-the_local_structure_of_poisson_manifolds}.
In order to be able to describe the action of the special linear group $\SLh$ by means of vector fields on $\mathfrak{T}_{1}(\Hh)$ we have to introduce a symmetric bivector field $\mathcal{R}$.
This bivector field is related to the Jordan algebra structure of $\uh$ coming from the anti-commutator product in $\mathcal{B}(\Hh)$.
Specifically, denoting by $f_{\mathbf{A}},f_{\mathbf{B}}$ the linear functions on $\mathfrak{T}_{1}(\Hh)$ associated with $\mathbf{A},\mathbf{B}\in\uh$, we have:

\be
\mathcal{R}(\mathrm{d}f_{\mathbf{A}},\,\mathrm{d}f_{\mathbf{B}}) := f_{\mathbf{A}\mathbf{B} + \mathbf{B}\mathbf{A}} - 2\,f_{\mathbf{A}}\,f_{\mathbf{B}}\,.
\ee
Since the differentials of the linear functions generate the cotangent space of $\mathfrak{T}_{1}(\Hh)$ at each point, the definition of $\mathcal{R}$ is well-posed.
With the help of the symmetric bivector we define the gradient-like vector fields $\widetilde{Y}^{\mathbf{A}}:=\mathcal{R}(\mathrm{d}f_{\mathbf{A}})$, and, according to the results in \cite{ciaglia_dicosmo_ibort_laudato_marmo-dynamical_vector_fields_on_the_manifold_of_quantum_states}, these vector fields together with the Hamiltonian ones $L^{\mathbf{A}}$ close on a realization of $\slh$ that integrates to the group action of $\SLh$ on $\stsp$ giving rise to the submanifolds of quantum states with fixed rank.
Consequently, we see that, by means of the bivector fields $\Lambda$ and $\mathcal{R}$, the group actions determining the two stratifications of the space of quantum states $\stsp$ are related with the Lie-Jordan algebra structure of $\uh$, which, in turns, is directly related with the $C^{*}$-algebra structure of $\mathcal{B}(\Hh)$.

In section \ref{sec: Kahler structures on the manifolds of isospectral quantum states} we focus on the submanifolds of isospectral quantum states, and review the K\"{a}hler structure it is possible to define on them.
This K\"{a}hler structure is invariant with respect to the action of $\SUh$, that is:

\be
\mathcal{L}_{L^{\mathbf{A}}}\,\omega = \mathcal{L}_{L^{\mathbf{A}}}\,g = \mathcal{L}_{L^{\mathbf{A}}}\,J = 0\,,\;\;\;\forall\,\mathbf{A}\in\suh\,,
\ee
where $\omega,\,g,\,J$ are, respectively, the symplectic form, the metric tensor, and the complex structure generating the K\"{a}hler structure.
The invariance property of the K\"{a}hler structure is the crucial ingredient we need to prove the main result given in section \ref{sec: Lie groups, Kahler actions, and complexification}.
Specifically, analogously with what is done for the gradient-like vector fields on $\mathfrak{T}_{1}(\Hh)$, we define gradient vector fields $Y_{f}$ on every submanifold of isospectral quantum states by means of the inverse $g^{-1}$ of the metric tensor $g$, namely, we set $Y_{f}:=(g^{-1})(\mathrm{d}f)$.
Because of the compatibility condition characterizing the K\"{a}hler structure, we have $Y_{f}=J(X_{f})$, where $X_{f}$ is the Hamiltonian vector field associated with $f$ by means of the symplectic structure $\omega$.
The gradient vector fields associated with the fundamental vector fields $L^{\mathbf{A}}$ generating the action of $\SUh$ are denoted as $Y^{\mathbf{A}}$.
Note that, unlike the gradient-like vector fields $\widetilde{Y}^{\mathbf{A}}$, the gradient vector fields are tangent to the submanifolds of isospectral quantum states by construction.
This means that the one-parameter group of diffeomorphisms they generate preserves the spectrum of quantum states even if it is not a unitary transformation.
In particular, the flow of the gradient vector fields preserve the von-Neumann entropy.
Once we have the Hamiltonian and gradient vector fields $L^{\mathbf{A}}$ and $Y^{\mathbf{A}}$, we are able to prove that they close on a realization of the Lie algebra $\slh$ on every submanifold of isospectral quantum states, and, since these manifolds are compact, the realization of $\slh$ integrates to an action of $\SLh$.
Clearly, this action is completely different from the action of $\SLh$ on $\stsp$ giving rise to the submanifolds of quantum states with fixed rank because both Hamiltonian and gradient vector fields are tangent to the submanifolds of isospectral quantum states.
We actually extend this result to the more general case of a K\"{a}hler manifold $M$ on which there is a Lie group $G$ acting by means of isometries of the K\"{a}hler structure.
In this case, we prove that there is always a realization of the complexification $\mathfrak{g}^{\mathbb{C}}$ of the Lie algebra $\mathfrak{g}$ of $G$ in terms of Hamiltonian and gradient vector fields.
Therefore, the case in which $M$ is actually a manifold of isospectral quantum states is obtained as a particular case of a more general instance.
Always in section \ref{sec: Lie groups, Kahler actions, and complexification}, we give an ``extrinsic'' description of the gradient vector fields in terms of a $(1,1)$ tensor field $\mathscr{J}$ on $\traceh$ defined starting from the Poisson tensor $\Lambda$ and the Euclidean metric tensor $\mathcal{N}$.
This extrinsic point of view allows to exploit the global differential geometry of the affine space $\traceh$ when performing explicit computations.

In section \ref{sec: q-bit II} we concretely analyze the case of a two-level quantum system.
In this specific case, we are able to prove that it is possible to describe the isospectral action of $\SLh\equiv\mathcal{SL}(2,\mathbb{C})$ in terms of Hamiltonian and the so-called pseudo-gradient vector fields on $\mathfrak{T}_{1}(\Hh)$.
These vector fields are defined applying a $(1,1)$ tensor field $\mathfrak{J}$ to the Hamiltonian vector fields $L^{\mathbf{A}}$ generating the action of $\SUh$.
In order for the pseudo-gradient vector fields to have the correct properties for describing the isospectral action of $\SLh$, the $(1,1)$ tensor field $\mathfrak{J}$ must be invariant with respect to the action of $\SUh$, and it must be such that $\mathfrak{J}^{3}=-\mathfrak{J}$.
This last property allows us to think of $\mathfrak{J}$ as being a sort of generalization of a complex structure to the case of an odd-dimensional manifold.
Indeed, every complex structure $J$ is such that $J^{3}=-J$, but the converse is not true.
Exploiting $\mathfrak{J}$, we are able to build another symmetric bivector field $G=\mathfrak{J}\circ\Lambda$ on $\mathfrak{T}_{1}(\Hh)$.
This bivector field plays a role analogous to that played by the metric tensors on the K\"{a}hler submanifolds of isospectral quantum states.
Furthermore, we see that $\Lambda=\mathfrak{J}\circ\mathcal{R}$, from which we conclude that the two symmetric bivector fields $\mathcal{R}$ and $G$, which are responsible for the definition of the gradient-like and pseudo-gradient vector fields associated with the two different actions of $\SLh$ on $\stsp$, are generated by $\Lambda$ and $\mathfrak{J}$ alone.
Indeed, the fact that both $\Lambda$ and $\mathfrak{J}$ are not invertible implies that the relation $G=\mathfrak{J}\circ \Lambda$ can not be inverted in order to give the relation $\Lambda=\mathfrak{J}\circ\mathcal{R}$ which may be seen as defining $\mathcal{R}$ implicitely.
In some sense, we may look at $(\Lambda,\, \mathfrak{J},\,G,\,\mathcal{R})$ as a sort of (contravariant) generalization of a K\"{a}hler manifold to a possibly odd-dimensional manifold.
We plan to expand on this topic in a future work.

In section \ref{sec: composite systems} we briefly comment on the possible application of the group actions associated with the geometrical structure of the space of quantum states $\stsp$ to the case of multipartite systems.
In particular, we note that, for a bipartite quantum system modelled on $\mathcal{H}_{1}\otimes\mathcal{H}_{2}$, the local action of $\mathcal{SL}(\mathcal{H}_{1})\otimes\mathcal{SL}(\mathcal{H}_{2})$ preserve the Schmidt rank of a quantum state $\rho$.
This observation, although immediate, seems to be new in the literature and it opens up the possibility to analyze entanglement/separability properties of a multipartite quantum system from the point of view of a Lie group which is ``bigger'' than the local unitary group usually employed in the literature.
Furthermore, the action of $\SLh$ generating the submanifolds of quantum states with fixed rank seems to offer a fruitful way of generalizing the Lorentz-group approach to the separability properties of a quantum system made of  two q-bits exposed in  \cite{samuel_shivam_sinha-lorentzian_geometry_of_qubit_entanglement} to higher dimensional quantum systems.
We proceed in studying the separability properties of density states in relation to the action of $\SLh$ preserving the rank.
In particular, for a bipartite quantum system distinguishable ``particles'' for which $\Hh=\Hh_{A}\otimes\Hh_{B}$, we introduce the notion of $(A,B)$-rank for product states and prove that product states with fixed $(A,B)$-rank are homogeneous spaces for the local action of $\mathcal{SL}(\Hh_{A})\times\mathcal{SL}(\Hh_{B})$.
Specifically, the $(A,B)$-rank for a product state $\rho_{AB}=\rho_{A}\otimes\rho_{B}$ is defined as the couple of integer numbers $(\mathrm{rk}(\rho_{A}),\,\mathrm{rk}(\rho_{B}))$ with $\mathrm{rk}(\rho_{A})$ and $\mathrm{rk}(\rho_{B})$ being the rank of $\rho_{A}$ and $\rho_{B}$ as linear operators on $\Hh_{A}$ and $\Hh_{B}$ respectively.
Note that two product states with the same $(A,B)$-rank  must necessarily have the same rank as linear operators on $\Hh$, while product states with the same rank may have different $(A,B)$-rank.
Two special example of manifolds of product states with fixed $(A,B)$-rank are the manifolds of quantum states with $(1,1)$-rank and the manifolds of quantum states with $(n_{A},n_{B})$-rank where $n_{A}=\mathrm{dim}(\Hh_{A})$ and $n_{B}=\mathrm{dim}(\Hh_{B})$.
The former coincides with the manifold of separable pure quantum states, while the latter is the manifolds of invertible product states.
The way in which these results are proved is such that they can immediately be extended to the case of a multipartite system of distinguishable ``particles''.
Finally, the relation between the separability properties of isospectral quantum states and the isospectral action of $\SLh$ introduced in section \ref{sec: Lie groups, Kahler actions, and complexification} is yet to be studied, and we plan to take on this task in future works.

\section*{Acknowledgements}

D.C. was supported by the National Science
Centre project 2015/19/B/ST1/03095.
A.I. and G.M. acknowledge financial support from the Spanish Ministry of Economy and Competitiveness, through the Severo Ochoa Programme for Centres of Excellence in RD (SEV-2015/0554).
A.I., would like to thank partial support provided by the MINECO research project MTM2014-54692-P and QUITEMAD+, S2013/ICE-2801.   
G.M. would like to thank the support provided by the Santander/UC3M Excellence Chair Programme 2016/2017.

%

\addcontentsline{toc}{section}{References}
\begin{thebibliography}{10}

\bibitem{abraham_marsden-foundations_of_mechanics}
R.~Abraham and J.~E. Marsden.
\newblock {\em {Foundations of Mechanics}}.
\newblock Addison-Wesley, Menlo Park, CA, second edition, 1978.

\bibitem{abraham_marsden_ratiu-manifolds_tensor_analysis_and_applications}
R.~Abraham, J.~E. Marsden, and T.~Ratiu.
\newblock {\em {Manifolds, tensor analysis, and applications}}.
\newblock Springer-Verlag, New York, 3rd edition, 2012.

\bibitem{aniello_clemente-gallardo_marmo_volkert-classical_tensors_and_quantum_entanglement_I}
P.~Aniello, J.~Clemente-Gallardo, G.~Marmo, and G.~F. Volkert.
\newblock {Classical Tensors and Quantum Entanglement I: Pure States}.
\newblock {\em International Journal of Geometric Methods in Modern Physics},
  7(3):485 -- 503, 2010.

\bibitem{ashtekar_schilling-geometrical_formulation_of_quantum_mechanics}
A.~Ashtekar and T.~A. Schilling.
\newblock Geometrical formulation of quantum mechanics.
\newblock In A.~Harvey, editor, {\em On Einstein's Path: Essays in Honor of
  Engelbert Schucking}, pages 23 -- 65. Springer-Verlag, New York, 1999.

\bibitem{ay_tuschmann-duality_versus_flatness_in_quantum_information_geometry}
N.~Ay and W.~Tuschmann.
\newblock {Duality versus dual flatness in quantum information geometry}.
\newblock {\em Journal of Mathematical Physics}, 44(4):1512 -- 1518, 2003.

\bibitem{bengtsson_zyczkowski-geometry_of_quantum_states:_an_introduction_to_quantum_entanglement}
I.~Bengtsson and K.~\.Zyczkowski.
\newblock {\em {Geometry of Quantum States: An Introduction to Quantum
  Entanglement}}.
\newblock Cambridge University Press, New York, 2006.

\bibitem{cantoni-generalized_transition_probability}
V.~Cantoni.
\newblock {Generalized ``Transition Probability''}.
\newblock {\em Communications in Mathematical Physics}, 44:125 -- 128, 1975.

\bibitem{cantoni-the_riemannian_structure_on_the_space_of_quantum-like_systems}
V.~Cantoni.
\newblock {The Riemannian Structure on the States of Quantum-like Systems}.
\newblock {\em Communications in Mathematical Physics}, 56:189 -- 193, 1977.

\bibitem{carinena_clemente-gallardo_jover-galtier_marmo-tensorial_dynamics_on_the_space_of_quantum_states}
J.~F. Cari\~{n}ena, J.~Clemente-Gallardo, J.~A. Jover-Galtier, and G.~Marmo.
\newblock {Tensorial dynamics on the space of quantum states}.
\newblock {\em Journal of Physics A}, 50(36):365301--30, 2017.

\bibitem{carinena_clemente-gallardo_marmo-geometrization_of_quantum_mechanics}
J.~F. Cari\~{n}ena, J.~Clemente-Gallardo, and G.~Marmo.
\newblock {Geometrization of quantum mechanics}.
\newblock {\em Theoretical and Mathematical Physics}, 152(1):894 -- 903, 2007.

\bibitem{chirco_mele_oriti_vitale-fisher_metri_geometric_entanglement_and_spin_networks}
G.~Chirco, F.~M. Mele, D.~Oriti, and P.~Vitale.
\newblock {Fisher metric, geometric entanglement, and spin networks}.
\newblock {\em Physical Review D}, 97(2):046015--29, 2018.

\bibitem{chruscinski_jamiolkowski-geometric_phases_in_classical_and_quantum_mechanics}
D~Chru\'{s}ci\'{n}ski and A.~Jamio\l{l}kowski.
\newblock {\em Geometric Phases in Classical and Quantum Mechanics}.
\newblock Birkh\"{a}user, Boston, 2004.

\bibitem{chruscinski_marmo-remarks_on_the_gns_representation_and_the_geometry_of_quantum_states}
D.~Chru\'{s}ci\'{n}ski and G.~Marmo.
\newblock {Remarks on the GNS Representation and the Geometry of Quantum
  States}.
\newblock {\em Open Systems \& Information Dynamics}, 16(2 and 3):155 -- 177,
  2009.

\bibitem{chruscinski_pascazio-a_brief_history_of_the_gkls_equation}
D.~Chru\'{s}ci\'{n}ski and S.~Pascazio.
\newblock {A Brief History of the GKLS Equation}.
\newblock {\em Open Systems \& Information dynamics}, 24(3):1740001--20, 2017.

\bibitem{ciaglia_dicosmo_felice_mancini_marmo_perez-pardo-aspects_of_geodesical_motion_with_fisher-rao_metric:classical_and_quantum}
F.~M. Ciaglia, F.~Di~Cosmo, D.~Felice, S.~Mancini, G.~Marmo, and J.~M.
  P\'{e}rez-Pardo.
\newblock
  \href{https://www.worldscientific.com/doi/abs/10.1142/S1230161218500051}{Aspects
  of geodesical motion with Fisher-Rao metric: classical and quantum}.
\newblock {\em Open Systems \& Information dynamics}, 25(1):1850005--14, 2018.

\bibitem{ciaglia_dicosmo_ibort_laudato_marmo-dynamical_vector_fields_on_the_manifold_of_quantum_states}
F.~M. Ciaglia, F.~Di~Cosmo, A.~Ibort, M.~Laudato, and G.~Marmo.
\newblock
  \href{https://www.worldscientific.com/doi/abs/10.1142/S1230161217400030}{Dynamical
  vector fields on the manifold of quantum states}.
\newblock {\em Open Systems \& Information dynamics}, 24(3):1740003--38, 2017.

\bibitem{ciaglia_dicosmo_ibort__marmo-dynamical_aspects_in_the_quantizer-dequantizer_formalism}
F.~M. Ciaglia, F.~Di~Cosmo, A.~Ibort, and G.~Marmo.
\newblock
  \href{https://www.sciencedirect.com/science/article/pii/S0003491617302567?via}{Dynamical
  aspects in the quantizer-dequantizer formalism}.
\newblock {\em Annals of Physics}, 385:769 -- 781, 2017.

\bibitem{ciaglia_dicosmo_laudato_marmo-differential_calculus_on_manifolds_with_boundary.applications}
F.~M. Ciaglia, F.~Di~Cosmo, M.~Laudato, and G.~Marmo.
\newblock
  \href{https://www.worldscientific.com/doi/abs/10.1142/S0219887817400035}{Differential
  Calculus on Manifolds with Boundary. Applications}.
\newblock {\em International Journal of Geometrical Methods in Modern Physics},
  14(8):1740003--39, 2017.

\bibitem{ciaglia_dicosmo_laudato_marmo_mele_ventriglia_vitale-a_pedagogical_intrinsic_approach_to_relative_entropies_as_potential_functions_of_quantum_metrics}
F.~M. Ciaglia, F.~Di~Cosmo, M.~Laudato, G.~Marmo, G.~Mele, F.~Ventriglia, and
  P.~Vitale.
\newblock
  \href{https://www.sciencedirect.com/science/article/pii/S0003491618301519}{A
  Pedagogical Intrinsic Approach to Relative Entropies as Potential Functions
  of Quantum Metrics: the q-z family}.
\newblock {\em Annals of Physics}, 395:238 -- 274, 2018.

\bibitem{ciaglia_ibort_marmo-geometrical_structures_for_classical_and_quantum_probability_spaces}
F.~M. Ciaglia, A.~Ibort, and G.~Marmo.
\newblock
  \href{https://www.worldscientific.com/doi/abs/10.1142/S021974991740007X}{Geometrical
  structures for classical and quantum probability spaces}.
\newblock {\em International Journal of Quantum Information},
  15(8):1740007--14, 2017.

\bibitem{ciaglia_ibort_marmo-differential_geometry_of_quantum_states_observables_and_evolution}
F.~M. Ciaglia, A.~Ibort, and G.~Marmo.
\newblock {Differential Geometry of Quantum States, Observables and Evolution}.
\newblock {\em In preparation}, 2018.

\bibitem{cirelli_lanzavecchia_mania-normal_pure_states_and_the_von_neumann_algebra_of_bounded_operators_as_kahler_manifold}
R.~Cirelli, P.~Lanzavecchia, and A.~Mania.
\newblock {Normal pure states of the von Neumann algebra of bounded operators
  as K\"{a}hler manifold}.
\newblock {\em Journal of Physics A: Mathematical General}, 16(16):3829 --
  3835, 1983.

\bibitem{cirelli_mania_pizzocchero-quantum_mechanics_as_an_infinite_dimensional_Hamiltonian_system_with_uncertainty_structure}
R.~Cirelli, A.~Mania, and L.~Pizzocchero.
\newblock {Quantum mechanics as an infinite dimensional Hamiltonian system with
  uncertainty structure}.
\newblock {\em Journal of Mathematical Physics}, 31(12):2891 -- 2903, 1990.

\bibitem{cirelli_pizzocchero-on_the_integrability_of_quantum_mechanics_as_an_infinite_dimensional_system}
R.~Cirelli and L.~Pizzocchero.
\newblock {On the integrability of quantum mechanics as an infinite dimensional
  Hamiltonian system}.
\newblock {\em Nonlinearity}, 3(4):1057 -- 1080, 1990.

\bibitem{ercolessi_marmo_morandi-from_the_equations_of_motion_to_the_canonical_commutation_relations}
E.~Ercolessi, G.~Marmo, and G.~Morandi.
\newblock {From the equations of motion to the canonical commutation
  relations}.
\newblock {\em Rivista del Nuovo Cimento}, 33:401 -- 590, 2010.

\bibitem{ercolessi_marmo_morandi_mukunda-geometry_of_mixed_states_and_degenerancy_structure_of_geometric_phases_for_multi-level_quantum_systems_a_unitary_group_approach}
E.~Ercolessi, G.~Marmo, G.~Morandi, and N.~Mukunda.
\newblock {Geometry of mixed states and degenerancy structure of geometric
  phases for multi-level quantum systems. A unitary group approach}.
\newblock {\em International Journal of Modern Physics A}, 16(31):5007 -- 5032,
  2001.

\bibitem{facchi_kulkarni_manko_marmo_sudarshan_ventriglia-classical_and_quantum_fisher_information_in_the_geometrical_formulation_of_quantum_mechanics}
P.~Facchi, R.~Kulkarni, V.~I. Man'ko, G.~Marmo, E.~C.~G. Sudarshan, and
  F.~Ventriglia.
\newblock {Classical and quantum Fisher information in the geometrical
  formulation of quantum mechanics}.
\newblock {\em Physics Letters A}, 374(48):4801 -- 4803, 2010.

\bibitem{gabach-clement_raggio-a_simple_spectral_condition_implying_separability_for_states_of_bipartite_quantum_systems}
M.~E. Gabach~Cl\'{e}ment and G.~A. Raggio.
\newblock {A simple spectral condition implying separability for states of
  bipartite quantum systems}.
\newblock {\em Journal of Physics A: Mathematical and General}, 39(29), 2006.

\bibitem{grabowski_kus_marmo-geometry_of_quantum_systems_density_states_and_entanglement}
J.~Grabowski, M.~Ku{\'s}, and G.~Marmo.
\newblock {Geometry of quantum systems: density states and entanglement}.
\newblock {\em Journal of Physics A: Mathematical and General}, 38(47):10217 --
  10244, 2005.

\bibitem{grabowski_kus_marmo-symmetries_group_actions_and_entanglement}
J.~Grabowski, M.~Ku{\'s}, and G.~Marmo.
\newblock {Symmetries, group actions, and entanglement}.
\newblock {\em Open Systems \& Information Dynamics}, 13(04):343 -- 362, 2006.

\bibitem{grabowski_kus_marmo-on_the_relation_between_states_and_maps_in_infinite_dimensions}
J.~Grabowski, M.~Ku{\'s}, and G.~Marmo.
\newblock {On the relation between states and maps in infinite dimensions}.
\newblock {\em Open Systems \& Information Dynamics}, 14(4):355 -- 370, 2007.

\bibitem{gurvits_barnum-separable_balls_around_the_maximally_mixed_multipartite_quantum_state}
L.~Gurvits and H.~Barnum.
\newblock {Separable balls around the maximally mixed multipartite quantum
  state}.
\newblock {\em Physical Review A}, 68(4), 2003.

\bibitem{horodecki-quantum_entanglement}
R.~Horodecki, P.~Horodecki, M.~Horodecki, and K.~Horodecki.
\newblock {Quantum entanglement}.
\newblock {\em Review of Modern Physics}, 81(2):865 -- 942, 2009.

\bibitem{huckleberry_kus_sawicki-bipartite_entanglement_spherical_ations_and_geometry_of_local_unitary_orbits}
A.~Huckleberry, M.~Ku{\'s}, and A.~Sawicki.
\newblock {Bipartite entanglement, spherical actions, and geometry of local
  unitary orbits }.
\newblock {\em Journal of Mathematical Physics}, 54(2):022202--19, 2013.

\bibitem{kibble-geometrization_of_quantum_mechanics}
T.~W.~B. Kibble.
\newblock {Geometrization of Quantum Mechanics}.
\newblock {\em Communications in Mathematical Physics}, 65(2):189 -- 201, 1979.

\bibitem{laudato_marmo_mele_ventriglia_vitale-tomographic_reconstruction_of_quantum_metrics}
M.~Laudato, G.~Marmo, G.~Mele, F.~Ventriglia, and P.~Vitale.
\newblock {Tomographic reconstruction of quantum metrics}.
\newblock {\em Journal of Physics A: Mathematical and Theoretical},
  51(5):055302--17, 2018.

\bibitem{leinaas_myrheim_ovrum-geometrical_aspects_of_entanglement}
J.~M. Leinaas, J.~Myrheim, and E.~Ovrum.
\newblock {Geometrical aspects of entanglement}.
\newblock {\em Physical Review A}, 74(1), 2006.

\bibitem{mackenzie-general_theory_of_lie_groupoids_and_algebroids}
K.~C. Mackenzie.
\newblock {\em General theory of Lie groupoids and Lie algebroids}.
\newblock Cambridge University Press, 2005.

\bibitem{manko_marmo_ventriglia_vitale-metric_on_the_space_of-quantum_states_from_relative_entropy_tomographic_reconstruction}
V.~I. Man'ko, G.~Marmo, F.~Ventriglia, and P.~Vitale.
\newblock {Metric on the space of quantum states from relative entropy.
  Tomographic reconstruction}.
\newblock {\em Journal of Physics A: Mathematical and Theorerical}, 50:302 --
  335, 2016.

\bibitem{marmo_scolarici_simoni_ventriglia-the_quantum-classical_transition:the_fate_of_the_complex_structure}
G.~Marmo, G.~Scolarici, A.~Simoni, and F.~Ventriglia.
\newblock {The quantum-classical transition: the fate of the complex
  structure}.
\newblock {\em International Journal of Geometrical Methods in Modern Physics},
  2:127 -- 145, 2005.

\bibitem{marsden_ratiu-introduction_to_mechanics_and_symmetry}
J.~E. Marsden and T.~Ratiu.
\newblock {\em Introduction to Mechanics and Symmetry}.
\newblock Springer-Verlag New York, 2nd edition, 1999.

\bibitem{marmo_ferrario_lovecchio_morandi_rubano-the_inverse_problem_in_the_calculus_of_variations_and_the_geometry_of_the_tangent_bundle}
G.~Morandi, C.~Ferrario, G.~Lo~Vecchio, G.~Marmo, and C.~Rubano.
\newblock {The inverse problem in the calculus of variations and the geometry
  of the tangent bundle}.
\newblock {\em Physics Reptorts}, 188(3 and 4):147 -- 284, 1990.

\bibitem{cencov_morozowa-markov_invariant_geometry_on_state_manifolds}
E.~A. Morozowa and N.~N. Cencov.
\newblock {Markov invariant geometry on state manifolds}.
\newblock {\em Journal of Soviet Mathematics}, 56(5):2648 -- 2669, 1991.

\bibitem{nirenberg_newlander-complex_analytic_coordinates_in_almost_complex_manifolds}
L.~Nirenberg and A.~Newlander.
\newblock {Complex Analytic Coordinates in Almost Complex Manifolds}.
\newblock {\em Annals of Mathematics}, 65(3):391 -- 404, 1957.

\bibitem{petz-monotone_metrics_on_matrix_spaces}
D.~Petz.
\newblock {Monotone metrics on matrix spaces}.
\newblock {\em Linear Algebra and its Applications}, 244:81 -- 96, 1996.

\bibitem{samuel_shivam_sinha-lorentzian_geometry_of_qubit_entanglement}
J.~Samuel, K.~Shivam, and S.~Sinha.
\newblock {Lorentzian geometry of qubit entanglement}.
\newblock {\em arXiv:1801.00611 [quant-ph]}, 2018.

\bibitem{sanpera_bruss_lewenstein-schmidt-number_witnesses_and_bound_entanglement}
A.~Sanpera, D.~Bruss, and M.~Lewenstein.
\newblock {Schmidt-number witnesses and bound entanglement}.
\newblock {\em Physical Review A}, 63(5), 2001.

\bibitem{huckleberry_kus_sawicki-symplectic_geometry_of_entanglement}
A.~Sawicki, A.~Huckleberry, and M.~Ku{\'s}.
\newblock {Symplectic Geometry of Entanglement }.
\newblock {\em Communications in Mathematical Physics}, 305:441 -- 468, 2011.

\bibitem{kus_oszmaniec_sawicki-convexity_of_momentum_map_morse_index_and_quantum_entanglement}
A.~Sawicki, M.~Oszmaniec, and M.~Ku{\'s}.
\newblock {Convexity of momentum map, Morse index, and quantum entanglement}.
\newblock {\em Reviews in Mathematical Physics}, 26(3):1450004--39, 2014.

\bibitem{terhal_horodecki-schmidt_number_for_density_matrices}
B.~Terhal and P.~Horodecki.
\newblock {Schmidt number for density matrices}.
\newblock {\em Physical Reviews A}, 61(4), 2000.

\bibitem{verstraete_audenaert_demoor-maximally_entangled_mixed_states_of_two_qubits}
F.~Verstraete, K.~Audenaert, and B.~De~Moor.
\newblock {Maximally entangled mixed states of two qubits}.
\newblock {\em Physical Review A}, 64(1), 2001.

\bibitem{weinstein-the_local_structure_of_poisson_manifolds}
A.~Weinstein.
\newblock {The local structure of Poisson manifolds}.
\newblock {\em Journal of Differential Geometry}, 18:523 -- 557, 1983.

\bibitem{weinstein-the_local_structure_of_poisson_manifolds_errata_and_addenda}
A.~Weinstein.
\newblock {The local structure of Poisson manifolds: errata and addenda}.
\newblock {\em Journal of Differential Geometry}, 22:255 -- 255, 1985.

\bibitem{wootters-statistical_distance_and_hilbert_space}
W.~K. Wootters.
\newblock Statistical distance and hilbert space.
\newblock {\em Physical Review D}, 23(2):357 -- 362, 1981.

\bibitem{zyczkowski_horodecki_sanpera_lewenstein-volume_of_the_set_of_separable_states}
K.~\.Zyczkowski, P.~Horodecki, A.~Sanpera, and M.~Lewenstein.
\newblock {Volume of the set of separable states}.
\newblock {\em Physical Review A}, 58(2), 1998.

\end{thebibliography}

\end{document}